\documentclass[preprint,secnumarabic,showpacs,amsmath,amssymb,floatfix,superscriptaddress,longbibliography,aps]{revtex4-1}

\usepackage{graphicx,color}
\usepackage{bm,changes}

\begin{document}

\newcommand{\cp}[1]{{\color{red} #1}}

\newcommand{\mfs}{Mn$_{1-x}$Fe$_{x}$Si}
\newcommand{\mcs}{Mn$_{1-x}$Co$_{x}$Si}
\newcommand{\fcs}{Fe$_{1-x}$Co$_{x}$Si}
\newcommand{\cso}{Cu$_{2}$OSeO$_{3}$}

\newcommand{\rxx}{$\rho_{xx}$}
\newcommand{\rxy}{$\rho_{xy}$}
\newcommand{\rxytop}{$\rho_{\rm xy}^{\rm top}$}
\newcommand{\Drxyt}{$\Delta\rho_{\rm xy}^{\rm top}$}
\newcommand{\Sxy}{$\sigma_{xy}$}
\newcommand{\Sxya}{$\sigma_{xy}^A$}

\newcommand{\bco}{$B_{\rm c1}$}
\newcommand{\bct}{$B_{\rm c2}$}
\newcommand{\bao}{$B_{\rm A1}$}
\newcommand{\bat}{$B_{\rm A2}$}
\newcommand{\beff}{$B_{\rm eff}$}
\newcommand{\bmm}{$B_{\rm m}$}
\newcommand{\bs}{$B_{\rm S}$}
\newcommand{\btop}{$B_{\rm top}$}
\newcommand{\bnfl}{$B_{\rm NFL}$}

\newcommand{\tc}{$T_{\rm c}$}
\newcommand{\tmax}{$T_{\rm max}$}
\newcommand{\ts}{$T_{\rm S}$}

\newcommand{\pc}{$p_{\rm c}$}

\newcommand{\mb}{$\mu_0\,M/B$}
\newcommand{\dmdb}{$\mu_0\,\mathrm{d}M/\mathrm{d}B$}
\newcommand{\ddmddb}{$\mathrm{\mu_0\Delta}M/\mathrm{\Delta}B$}
\newcommand{\cm}{$\chi_{\rm M}$}
\newcommand{\cac}{$\chi_{\rm ac}$}
\newcommand{\rechi}{${\rm Re}\,\chi_{\rm ac}$}
\newcommand{\imchi}{${\rm Im}\,\chi_{\rm ac}$}

\newcommand{\ozz}{$\langle100\rangle$}
\newcommand{\ooz}{$\langle110\rangle$}
\newcommand{\ooo}{$\langle111\rangle$}
\newcommand{\too}{$\langle211\rangle$}

%%% Added by DA to highlight changes.
\newcommand{\change}[2]{{\color{red}\sout{#1}}{\color{blue}#2}}

\renewcommand{\floatpagefraction}{0.5}
\bibliographystyle{nature}

%%%%%%%%%%%%%%%%%%%%%%%%%%%%%%%%%%%

\renewcommand{\vec}[1]{{\bf #1}}
\newcommand{\eref}[1]{(\ref{#1})}
\newcommand{\uv}[1]{{\bf \hat #1}}
\newcommand{\tr}{\mathrm{tr}}
\renewcommand{\Re}{\mathrm Re}
\newcommand{\be}{\begin{equation}}
\newcommand{\ee}{\end{equation}}

%%%%%%%%%%%%%%%%%%%%%%%%%%%%%%%%%%%

\title{Observation of two independent skyrmion phases in a chiral magnetic material}

\author{A. Chacon}
\affiliation{Physik Department, Technische Universit\"at M\"unchen, D-85748 Garching, Germany}

\author{L. Heinen}
\affiliation{Institut f\"ur Theoretische Physik, Universit\"at zu K\"oln, D-50937 K\"oln, Germany}

\author{M. Halder}
\affiliation{Physik Department, Technische Universit\"at M\"unchen, D-85748 Garching, Germany}

\author{A. Bauer}
\affiliation{Physik Department, Technische Universit\"at M\"unchen, D-85748 Garching, Germany}

\author{W. Simeth}
\affiliation{Physik Department, Technische Universit\"at M\"unchen, D-85748 Garching, Germany}

\author{S. M\"uhlbauer}
\affiliation{Heinz Maier-Leibnitz Zentrum (MLZ), Technische Universit\"at M\"unchen, D-85748 Garching, Germany}

\author{H. Berger}
\affiliation{\'Ecole Polytechnique Federale de Lausanne, CH-1015 Lausanne, Switzerland}

\author{M. Garst}
\affiliation{Institut f\"ur Theoretische Physik, Technische Universit\"at Dresden, D-01062 Dresden, Germany}

\author{A. Rosch}
\affiliation{Institut f\"ur Theoretische Physik, Universit\"at zu K\"oln, D-50937 K\"oln, Germany}

\author{C. Pfleiderer}
\affiliation{Physik Department, Technische Universit\"at M\"unchen, D-85748 Garching, Germany}

\date{\today}

\maketitle

\textbf{
Magnetic materials can host skyrmions, which are topologically non-trivial spin textures. In chiral magnets with cubic lattice symmetry, all previously-observed skyrmion phases require thermal fluctuations to become thermodynamically stable in bulk materials, and therefore exist only at relatively high temperature, close to the helimagnetic transition temperature. Other stabilization mechanisms require a lowering of the cubic crystal symmetry. Here, we report the identification of a second skyrmion phase in {\cso} at low temperature and in the presence of an applied magnetic field. The new skyrmion phase is thermodynamically disconnected from the well-known, nearly-isotropic, high-temperature phase, and exists, in contrast, when the external magnetic field is oriented along the {\ozz} crystal axis only. Theoretical modelling provides evidence that the stabilization mechanism is given by well-known cubic anisotropy terms, and accounts for an additional observation of metastable helices tilted away from the applied field. The identification of two distinct skyrmion phases in the same material and the generic character of the underlying mechanism suggest a new avenue for the discovery, design, and manipulation of topological spin textures.
}

Historically conceived to account for nucleons as excitations of pion fields \cite{1962:Skyrme:NP}, the notion of skyrmions has attracted great interest for many decades in the context of nonlinear continuum models in condensed matter systems describing, for example, excitations of $^3$He, spin textures in quantum Hall magnets, or the blue phases in liquid crystals \cite{Brown:2010, Volovik, 2001:Khawaya:Nature}. In recent years, the discovery of skyrmions in magnetic materials has also generated great interest \cite{2009:Muhlbauer:Science, 2010:Munzer:PhysRevB,  2011:Yu:NatureMater, 2012:Seki:Science, 2013:Nagaosa:NN}. As a key ingredient, skyrmion phases require chiral spin-orbit interactions that originate from a lack of inversion symmetry. Also known as Dzyaloshinsky-Moriya interactions, they twist the magnetization on scales much larger than atomic distances \cite{Landau}. 

For materials that support, in decreasing order of strength, exchange interactions, spin-orbit coupling, and magnetic anisotropies, it has long been known that the mean-field energy of a possible skyrmion phase is only slightly larger than topologically trivial phases. Moreover, theoretical calculations suggested that easy-plane anisotropies could, in principle, reduce this energy difference under applied fields such that a skyrmion lattice stabilizes \cite{1989:Bogdanov}. In reciprocal space the resulting skyrmion lattice may then be approximated as a superposition of three helical modulations with wave vector $\vec k$, forming a planar triple-$\vec k$ state. 

Therefore, it was a major surprise when a skyrmion lattice was first identified experimentally in a small applied field near the helimagnetic to paramagnetic transition in magnetic materials with only very weak cubic anisotropies, notably MnSi \cite{2009:Muhlbauer:Science} and {\fcs} \cite{2010:Munzer:PhysRevB}, later followed by FeGe \cite{2011:Yu:NatureMater}, {\cso} \cite{2012:Seki:Science}, and the $\beta$-Mn series \cite{2015:Tokunaga:NatCommun}. Quantitative calculations established that the skyrmion lattice in these materials is stabilized by thermal fluctuations akin to the notion of order-by-disorder \cite{2009:Muhlbauer:Science, 2013:Buhrandt:PRB}. In turn, the temperature versus field regime of the skyrmion phase is almost isotropic for different field directions with respect to the crystal lattice \cite{Bauer_book_16}. 

This situation is contrasted by systems with lower crystal symmetry or stronger spin-orbit coupling, where magnetocrystalline anisotropies might favour topological magnetic textures--notably the rhombohedral lacunar spinels GaV$_8$S$_8$ and GaV$_8$Se$_8$ \cite{2015:Kezsmarki:NM,2017:Bordacs:SR}, the hexagonal M-type ferrites \cite{2011:Yu:PNAS}, and the tetragonal Heusler compounds \cite{2017:Nayak:Nature}. Similarly, in thin films and tailored heterostructures, the formation of skyrmions is aided by the influence of surface energies \cite{wilson:PRB:14}. The cubic material MnGe, which exhibits strong spin-orbit coupling, realizes a three-dimensional hedgehog texture, although the precise mechanism of the stabilization is unresolved \cite{Kanazawa:2016fd}. In all known cases, the equilibrium phase diagram hosts only a single skyrmion phase. Non-thermal control parameters such as strain or electric fields \cite{2015:Chacon:PRL, 2016:Okamura:NatComm} can induce shifts of the phase boundaries, and variations in the metastable properties under super-cooling have been observed \cite{2010:Munzer:PhysRevB, 2016:Oike:NaturePhys, 2016:Karube:NatMater}. However, the expectation that new stabilization mechanisms may lead to the emergence of independent additional skyrmion phases has not been fulfilled.

Here, we revisited the role of well-known, standard contributions to cubic magnetic anisotropies (see, for example, Ref.\,\cite{2015:Grigoriev:PRB}) and find such an independent second skyrmion phase. Using small-angle neutron scattering (SANS) (described in the Methods and Supplementary Figs.\,8 and 9), we systematically track the magnetic order in {\cso} over a wide parameter range, focussing on magnetic fields parallel to the $\langle 100 \rangle$ crystal axis. This regime has been addressed in previous studies, where unexplained hysteretic effects were seen \cite{2012:Belesi:PRB,2016:Qian:PRB}.  In a recent preprint, observation of the tilted conical phase was reported \cite{pappas}, however, the  low-temperature skyrmion phase representing the main message of our paper was neither observed experimentally nor anticipated theoretically.

In Fig.\,\ref{figure1} our main results are summarized. We find a highly hysteretic, thermodynamically stable skyrmion phase at low temperatures at the border between the conical and the spin-polarized state, for fields along the $\langle 100 \rangle$ axis. The new phase emerges in addition to the well-known, nearly isotropic skyrmion phase at high temperature in the same material. It is accompanied by a regime in which the conical modulation increasingly tilts away from the field. As the main new insight, excellent agreement with theoretical modelling demonstrates that well-known magnetocrystalline cubic anisotropies, being nonlinear in the magnetization squared, can stabilize topologically non-trivial phases in the presence of a finite magnetic field. This greatly increases the potential number of materials hosting topological spin textures. We also find that stabilization through cubic magnetocrystalline anisotropies may happen in addition to other mechanisms, providing access to an interplay of different forms of topological spin textures in the same material.

The intensity distributions we observe in the SANS data may be fully accounted for by the five contributions depicted in Figs.\,\ref{figure1}A. The scattering planes containing the defining characteristics are displayed in grey shading, where it is essential to note the field orientation with respect to these planes. Figure\,\ref{figure1}A1--\ref{figure1}A3 represent the well-established intensity distributions of the helical state along the $\langle100\rangle$ directions, the conical state (modulation parallel to the field), and the high-temperature skyrmion phase near $T_c$ (sixfold pattern perpendicular to field). The two new key characteristics that emerge deep within the conical state for fields along $\langle100\rangle$ are illustrated in Figs.\,\ref{figure1}A4, \ref{figure1}A5, comprising a tilting of the conical modulations away from the field direction by an angle $\theta$ and a pronounced ring of scattering intensity perpendicular to the applied field, respectively. Further experimental examination demonstrates that the ring of scattering represents a signature arising from the well-known six-fold skyrmion lattice scattering pattern as the true ground state. In the following we identify this ring of scattering as the signature of a second, thermodynamically independent skyrmion phase, accompanied by a metastable phase of conical helices tilted away from the field by an angle $\theta$.

Three temperature versus field protocols were carefully followed, as summarized in Figs.\,\ref{figure1}C--\ref{figure1}E, namely zero-field-cooled/field-heated (ZFC/FH), field-cooled (FC) and high-field-cooled/field-heated (HFC/FH) (see Methods and Supplementary Fig.\,11). Our main observations are as follows: the new phases we observe are highly hysteretic; the tilted conical intensity and the ring of intensity emerge for all protocols; the intensities of the tilted conical state and the ring increase strongly with decreasing temperature; the regimes of the two new features are disconnected from the high-temperature skyrmion phase, including metastable remnants of the high-temperature skyrmion phase under FC [see Fig.\,\ref{figure1}D]. (v); for all protocols the emergence of the new features commences with a tilting of the conical state, followed by the formation of the ring of intensity; once the ring of intensity emerged, it only vanished as a function of field or temperature after the intensities of the tilted conical state disappeared, that is, we never observed a transition from the phase characterized by the ring towards the tilted conical phase; small oscillations of the field amplitude increased the intensity of the ring and under moderate tilting of the field direction away from $\langle 100\rangle$ the ring developed a six-fold-symmetric intensity pattern.

Typical data observed under HFC/FH are shown in Fig.\,\ref{figure2} (for data under ZFC/FH and FC see Supplementary Figs.\,12 and 13 in SI). Figures \ref{figure2}A1 and \ref{figure2}B1 show typical integrated intensity patterns of rocking scans for fields parallel and perpendicular to the neutron beam, respectively. The intensities in sectors 1 and 2 in Fig.\,\ref{figure2}A1 correspond to helical and skyrmion states, whereas sectors 1 and 2 in Fig.\,\ref{figure2}B1 correspond to conical and tilted conical states. Intensity maps as determined for those sectors are shown in Figs.\,\ref{figure2}A2, \ref{figure2}A3, \ref{figure2}B2,\ref{figure2}B3 (tiny black dots designate temperatures and fields at which data were recorded). Helical order emerges with decreasing field, as shown in Fig.\,\ref{figure2}A2. The location of skyrmion states is shown in Fig.\,\ref{figure2}A3, forming separate regimes at low and high temperatures. The latter is underscored by the temperature dependence at 25\,mT, shown in Fig.\,\ref{figure2}A4, featuring two maxima, where the pattern changes from a ring to a six-fold symmetry at the local minimum around $50\,{\rm K}$ between the two maxima. This unusual temperature and field dependence of the skyrmion intensity compares with the absence of intensity of the conical state at low temperatures and high fields, as shown in Fig.\,\ref{figure2}B2. Here the tilted conical intensity emerges instead, as shown in Fig.\,\ref{figure2}B3. The temperature dependence at 70\,mT, shown in Fig.\,\ref{figure2}B4 highlights the finite temperature range of the conical state.

As shown in Fig.\,\ref{figure3}A for $T\approx 5\,{\rm K}$, the angle $\theta$ between the orientation of the conical modulation and the applied field, defined in Fig.\,\ref{figure2}B1, is vanishingly small for low fields. In contrast, above approximately $40\,{\rm mT}$ we find spots corresponding to the modulation axis tilting away from the field approaching $\pm30^{\circ}$ at high fields. The field dependence of the intensities of these spots under slight misalignment is characteristic of a multi-domain state. Despite the large hysteresis in the scattering intensities and the phase boundaries (see Fig.\,\ref{figure1}), the tilting angle is insensitive to the protocol followed. Moreover, as shown in Fig.\,\ref{figure3}B, the modulus, $\vert \vec{k} \vert$, of the different magnetic states at $\sim 5\,{\rm K}$ exhibits a strong field dependence up to approximately $30\,\%$ under a perpendicular applied field, where four situations may be distinguished by the colour of the plotted points: (white) for the conical state parallel to the field $\vert \vec{k} \vert$ is unchanged; (ii, gray) For the tilted conical state $\vert \vec{k} \vert$ decreases above approximately $45\,{\rm mT}$, subsequently tracking the tilting angle qualitatively; (orange) For the super-cooled high-temperature skyrmion state $\vert \vec{k} \vert$ decreases, no such decrease is observed for other crystallographic orientations (see Supplementary Information); (green and red) helical domains perpendicular to the field and skyrmion states at low temperatures display the same strong decrease of  $\vert \vec{k} \vert$, regardless of the protocol. 

When the sample was cooled in an applied field of 250\,mT along $\langle100\rangle$ down to 3.6\,K (HFC) and the field was reduced to $60\,{\rm mT}<\mu_0 H_{c2}\approx95{\rm mT}$, repeated cycling from 60\,mT to 80\,mT led to a successive increase of the intensity of the ring (LT-Sky) by over a factor of five, as shown in Fig.\,\ref{figure3}C. Thus, the ring of scattering, shown in Fig.\,\ref{figure3}D, is clearly connected with the ground state, but separated by a large energy barrier from the other phases.  Following this procedure, rotating the crystal at a field of 60\,mT with respect to the vertical axis by an angle $\phi$ (Supplementary Fig.\,9), while leaving the field direction unchanged, left the orientation of the ring of scattering unchanged perpendicular to the $\langle100\rangle$ axis. This is illustrated in Fig.\,\ref{figure3}E, which shows that a regular rocking scan (denoted $\omega$-scan) displays the same angular dependence as observed when rotating the $\langle100\rangle$ axis away from the field (denoted $\phi$-scan). At the same time $\vert \vec{k} \vert$ increased as shown in Fig.\,\ref{figure3}F, consistent with the projected field component along $\langle100\rangle$ (see Fig.\,\ref{figure3}B). Finally, as the rotation angle $\phi$ exceeded approximately $15^{\circ}$, the ring of scattering intensity developed a six-fold azimuthal dependence, which remained unchanged when returning the field direction back to $\langle100\rangle$ ($\phi=0$) as shown in Fig.\,\ref{figure3}G. This pattern is analogous to the skyrmion lattice at high temperatures. Its rocking width corresponds to that of the ring as shown in Fig.\,\ref{figure3}E, characteristic of an ordering process in the plane perpendicular to the field. 

Our main experimental observations that require theoretical explanation are the emergence of a phase characterized by a ring of scattering perpendicular to field and the existence of a tilted conical state. Small cyclic changes of the field amplitude connect the ground state with the ring of scattering, which develops a six-fold intensity variation, characteristic of a skyrmion lattice, under a rotation of the $\langle100\rangle$ axis away from the field direction.
The possible origin of these findings are strongly constrained in two ways. First, both features emerge with respect to a specific field direction only, namely $\langle 100\rangle$, so they must be due to cubic anisotropy terms. Second, the effective anisotropy is field-dependent: at $B=0$ helices prefer to align along the $\langle 100\rangle$ directions, but tilt away from this direction for fields parallel to $\langle 100\rangle$ once a threshold has been reached even though helices in the absence of magnetic anisotropies gain energy by aligning parallel to the field. Thus, the anisotropy must be increasing nonlinearly in the magnetization squared. 

For the $P2_13$ space group of {\cso} to lowest order in spin-orbit coupling only one anisotropy term allows to satisfy these criteria. It is given by
\begin{equation}\label{ani1}
F_a =- K \int d^3 {\bf r}\,(M_x^4+ M_y^4 + M_z^4)
\end{equation}
where $K$ is an anisotropy constant. This term represents the well-known textbook example of the leading-order cubic magnetocrystalline anisotropies in Fe and Ni, which has been found to influence the high-temperature skyrmion phase only very weakly. In the temperature and field range of interest its effects differ distinctly from the easy-plane anisotropies considered in earlier theoretical proposals \cite{1989:Bogdanov, wilson:PRB:14}, but compares with considerations in a recent experimental study \cite{2015:Grigoriev:PRB}. It is helpful to note that other anisotropy terms to the same order in spin-orbit coupling, such as $\alpha_1  M_x \partial_x^2 M_x+ \alpha_2 M_y \partial_x^2 M_y + \alpha_3 {\bf M} \partial_x^4 {\bf M}+{\rm cycl.}$, may be of similar strength, but cannot account for the increasing strength of the anisotropy under increasing field because they are quadratic in the magnetization. We also neglect more complicated anisotropy terms such as $\beta_1 M_x^6 + \beta_2  M_y \partial_x M_z^3 +{\rm cycl.}$, being formally weaker, since they contribute at higher order in the spin-orbit coupling. 

The main effects of Eq.\,\ref{ani1} under increasing magnetic fields along $\langle100\rangle$, in the following defined to be the $z$-axis, are due to contributions proportional to the square of the uniform magnetization $M_z^0$ that may be expressed analytically as
\begin{equation}
\Delta F_a=-6 K (M^z_0)^2 \sum_{{\bf k}\neq 0} M^z_{\bf{k}}M^z_{-\bf{k}}
\end{equation}
We determine the sign, $K>0$, from magnetization measurements of our samples (not shown) and similar measurements reported previously \cite{2016:Qian:PRB}. For this sign and vanishing magnetic field (and thus magnetization), helical order stabilizes along 
$\langle 100 \rangle$, consistent with experiment. For the conventional conical modulation with $\bf k$ parallel to $\hat z$ the $z$ component of the magnetization is not modulated, $M^z_{\bf{k}}=0$, and $\Delta F_a$ vanishes. In contrast, for both a skyrmion phase and a tilted conical phase, $M^z_{\bf{k}}$ is non-zero resulting in a reduction of the total energy linear in $K$, so that these two phases are stabilized by the anisotropy term. These considerations are corroborated by the decrease of $\theta$ and increase of $\vert \vec{k} \vert$ for increasing temperature (Supplementary Figs.\,18--22), as the reduction of the magnetization amplitude reduces the effects of anisotropy, which is experimentally found to vanish around 30\,K. 

To establish whether this mechanism allows the stabilization of a new ground state, we performed explicit numerical calculations where we minimized the free energy of an appropriate Ginzburg-Landau theory in momentum space (for technical details see Supplementary Section I). In addition to the anisotropy term, equation\,\ref{ani1}, our calculations took into account Zeeman energy, exchange, isotropic Dzyaloshinskii-Moriya, and dipolar interactions. The latter suppresses the spin-polarized phase in the regime where $\Delta F_a$ is sufficiently large, further reducing the energy of the tilted conical state and the additional skyrmion phases.

The calculated magnetic phase diagram is shown in Fig.~\ref{figure4}A, as inferred from the field dependence of the energy of the various magnetic states for a fixed value of $K$, which is shown in Fig.~\ref{figure4}\,B. Within the Ginzburg Landau approach, large negative values of $r_0$ correspond to temperatures far below the critical temperature $T_c$ ($r_0=-1$ corresponds to $T_c$). While for $K=0$ the conical phase is always the ground state, a skyrmion phase stabilizes when $K$ exceeds a threshold $K_c$ in a sufficiently large field at low temperatures, as expected from the qualitative arguments given above. 

In material-specific parameters, $K_c=K_{\sigma,c}\,/M_s^4$ may be expressed as the ratio $K_{\sigma,c}/(\mu_0 H^{\rm int}_{c2} M_s)= 0.07$, where $K_{\sigma,c}$ is the cubic anisotropy in units of energy density, $H^{\rm int}_{c2}$ is the critical field separating the conical from the field-polarized phase, and $M_s$ is the saturated magnetization  (roughly estimated $K_{\sigma,c}\approx400\,{\rm J\,m^{-3}}$). As this ratio scales like the square of spin-orbit coupling $K_{\sigma,c}/(\mu_0 H^{\rm int}_{c2} M_s) \sim \lambda_{\rm SOC}^2$, the new skyrmion phase is only stable if $\lambda_{\rm SOC}$ is sufficiently strong.

Further, using realistic quantitative values for the dipole-dipole interaction, we find for a small field range at the lowest temperatures (hatched area in Fig.\,\ref{figure4}A) that the tilted conical phase has a lower energy than both the conical and the ferromagnetic phase. However, for the parameters chosen, the skyrmion phase has always the lowest energy consistent with our experiments (Supplementary Fig.\,1). Our numerical results also exhibit an increase of $\theta$ with increasing field reaching roughly $30^{\circ}$ (Fig.~\ref{figure4}C), as well as a reduction of $\vert \vec{k} \vert$ by over $40\,\%$  (Fig.~\ref{figure4}\,D), in agreement with experiment. Qualitative differences, such as discontinuous changes of $\theta$ and $\vert \vec{k} \vert$, may be attributed to the omission of further subleading anisotropy terms which do not change our overall assessment.

For the parameters investigated in our calculations, the transition from the conical to the tilted conical state and the transition from the ferromagnetic to the tilted conical state appear to be second order and first order, respectively. In comparison, the transition to the skyrmion phase is strongly first order and additionally protected by topology. This accounts for the hysteretic behaviour we observe (Fig.~\ref{figure1}), where we expect a transition from the conical to the tilted conical state whenever the tilted conical state has a lower energy than the conical state, while the transition to the skyrmion phase involves a large energy barrier requiring the creation of pairs of Bloch points (or `magnetic monopoles')  \cite{2013:Milde:Science}. It is interesting to speculate whether domain walls of the tilted conical state may effectively decrease the energy barrier for skyrmion creation. Once the skyrmion phase has been generated, it is extremely robust, consistent with experiment. 

Regarding the morphology of the skyrmion phase we find that triangular and square lattices, as well as other distorted structures (not shown), have remarkably similar energies. This is consistent with experiment, where the ring of intensity shows that long-range crystalline magnetic order does not fully develop even though the skyrmion phase is thermodynamically stable. Instead, different patches of the skyrmion lattice display random orientations, as the in-plane anisotropy energy, determining the orientation of the skyrmion lattice, is much smaller than the energy barriers due to domain walls of the skyrmion lattice \cite{2017:Wild:SciAdv}. In fact, preliminary tests even suggest that the diffraction pattern that emerges from the ring under moderate crystal rotations, when recorded closer to $H_{c2}$, may support a four-fold pattern consistent with the predicted square skyrmion lattice.

The anisotropy term also explains why a tilt of the magnetic field is necessary to recover the six-fold symmetric scattering pattern characteristic of the skyrmion lattice. For a field in a $\langle100\rangle$ direction, $\phi=0$, the anisotropy term leads only to a very weak potential proportional to $-K^3 \cos(12 \gamma)$, where $\gamma$ is the angle determining the in-plane orientation of the skyrmion lattice. $K$ being small, this potential is apparently too weak to fix the rotational direction of skyrmion domains. The same is true for contributions arising from other anisotropy terms. However, in the presence of a magnetic field perpendicular to $\langle100\rangle$, a new anisotropy term, linear in $K$, is generated proportional to $K (B_{\perp} M_s)^2 \cos[6 (\gamma- \gamma_0/3)]$, with a six-fold symmetry. Here $\gamma_0$ parametrizes the orientation of $B_{\perp}$ in the $xy$ plane. This potential already dominates at small fields, as the prefactor is linear in $K$. Numerically, we find that the field-induced anisotropy term is enhanced by a numerical prefactor of order 10 compared to the $K^3$ term. The very strong hysteresis observed experimentally when rotating the $\langle100\rangle$ axis away from the field direction, illustrated in Fig.\,\ref{figure3}, maintains the alignment of the low-temperature skyrmion phase perpendicular to $\langle100\rangle$. In turn, the rotation of the crystal axis against the field leads to an in-plane component of the applied field, $B_{\perp}$, while the field component stabilising the skyrmion phase decreases. This generates the observed increase of $\vert \vec{k} \vert$ and the six-fold pattern as aligned along one of the in-plane $\langle 100\rangle$ axes.

Taken together, our findings indicate the discovery of a second skyrmion phase in {\cso} at low temperature, accompanied by a metastable tilted conical state, that is thermodynamically disconnected from the well-known skyrmion phase at high temperatures. To the best of our knowledge this represents the first case of two disconnected skyrmion phases in any material system, including bulk compounds, thin films, heterostructures, surfaces and nano-scaled devices. We show that the well-known cubic anisotropy term can strongly modify modulated magnetic structures when an external magnetic field is applied. This strongly motivates the reconsideration of a wide range of materials. While the magnitude of $K$ in MnSi, {\fcs}, and FeGe appears to be too small to reach this limit, it is interesting to speculate whether metastable properties in the $\beta$-Mn series \cite{2016:Karube:NatMater} arise from a second skyrmion phase in the presence of very strong disorder. 

The engineering of easy-axis or easy-plane anisotropies, described by quadratic terms in the free energy, already plays a decisive role for many experiments in multilayer magnetic materials. Much less attention has been given to engineer anisotropies represented by quartic terms such as the cubic anisotropies. For example, one could study their effects in materials where the quadratic anisotropy terms have been tuned to zero. In addition, it will be interesting to explore whether new inhomogeneous phases (supported by, for example, surface Dzyaloshinsky-Moriya interactions) can be controlled using effects similar to the ones discussed in this paper. Our study also reveals that two independent stabilization mechanisms may establish independent skyrmion phases in the same material. The new mechanism we identify is rather generic, and suggests that a wide range of magnetic materials with both cubic and lower-symmetry crystal structures may feature several disconnected skyrmion phases. The associated phases may, in principle, display different topological characteristics and morphologies in the same material, which, when tuned with additional parameters such as strain or electric field, offer a powerful new approach for applications.

%%%%%%%%%%%%%%%%%%%%%%%%%%%%%%%%%%
\newpage
\section*{Methods}

\textbf{Neutron scattering:} 
To explore the microscopic nature of the phase diagram of {\cso} we performed small angle neutron scattering (SANS) at the beam-line SANS-1 at FRM II (Munich) \cite{muhlbauer:NIaMiPRSAASDaAE:16}. Most data were recorded at temperatures down to 3.6\,K using a pulse tube cooler. Selected additional measurements using a $^3$He system confirmed the same behaviour down to 0.5\,K. Measurements were performed on a single crystal that was carefully polished into a sphere to eliminate inhomogeneities of the internal magnetic field due to demagnetising effects. The orientation of the crystal was determined by x-ray and neutron diffraction. The phase diagrams were determined for field along different crystallographic orientations combining identical temperature versus field histories for two scattering geometries: notably magnetic field parallel and perpendicular to the incident neutron beam, respectively. 

The three temperature versus field protocols used in our study may be summarised as follows. First, after zero field cooling (ZFC) down to the lowest temperature accessible in our study, the field was increased to the value of interest and data were recorded while successively heating (field-heating: FH). Second, data were recorded while cooling successively in a fixed applied field (field-cooling: FC) starting at $\sim70\,{\rm K}$. Third, after cooling the sample in the field-polarized state under $H>H_{c2}$ (high-field-cooling: HFC), and reducing the field to the value of interest, data were recorded while increasing the temperature (FH). 

\textbf{Theoretical modelling:} 
The theoretical calculations were based on a classical Ginzburg-Landau field theory. The mean-field phase diagram was calculated by minimizing an energy functional that included exchange, Dzyaloshinsky-Moriya, Zeeman, demagnetization, Ginzburg-Landau and the $M_x^4$-anisotropy energies. For explicit calculations the energy was discretized in momentum space before using a quasi-Newton minimization. To confirm the nature of the different magnetic phases the spin arrangement in real space as well as the winding number were calculated for selected parameter combinations.  For further details see SI.

\textbf{Data availability statement:} 
The data that support the plots within this paper and other findings of this study are available from the corresponding author upon reasonable request.

%%%%%%%%%%%%%%%%%%%%%%%%%%%%%%%%%%
\newpage
\section*{Acknowledgements}

We wish to thank S. Mayr, M. Meven and the team at FRM II for helpful discussions and support. AC, MH, and WS acknowledge financial support through the TUM Graduate School. This project has received funding from the European Research Council (ERC) under the European Union's Horizon 2020 research and innovation programme (grant agreement No 788031). LH and AR acknowledge financial support through DFG CRC1238 (project C02). M.G. acknowledges financial support from DFG CRC 1143 and DFG Grant 1072/5. AB, MH, WS, and CP acknowledge support through DFG TRR80 (projects E1, F2 and F7) as well as ERC-AdG (291079 TOPFIT). 

\section*{Author Contributions}
AC, MH, AB, WS, and SM performed the experimental work; 
AC analysed the data;
CP supervised the experimental work; 
HB grew the single crystals;
LH, MG, and AR developed the theoretical analysis;
AC and CP proposed this study and wrote the manuscript; 
all authors discussed the data and commented on the manuscript;
correspondence may be addressed to AC and CP.
\\

%%%%%%%%%%%%%%%%%%%%%%%%%%%%%%%%%%%
%%% Commented out by DA to be replaced by input from bbl file
%\newpage
%\section*{References}
%\bibliography{LT-Skx}

\clearpage
%%%%%%%%%%%%%%%%%%%%%%%%%%%
\centerline{\includegraphics[width=0.6\textwidth,clip=]{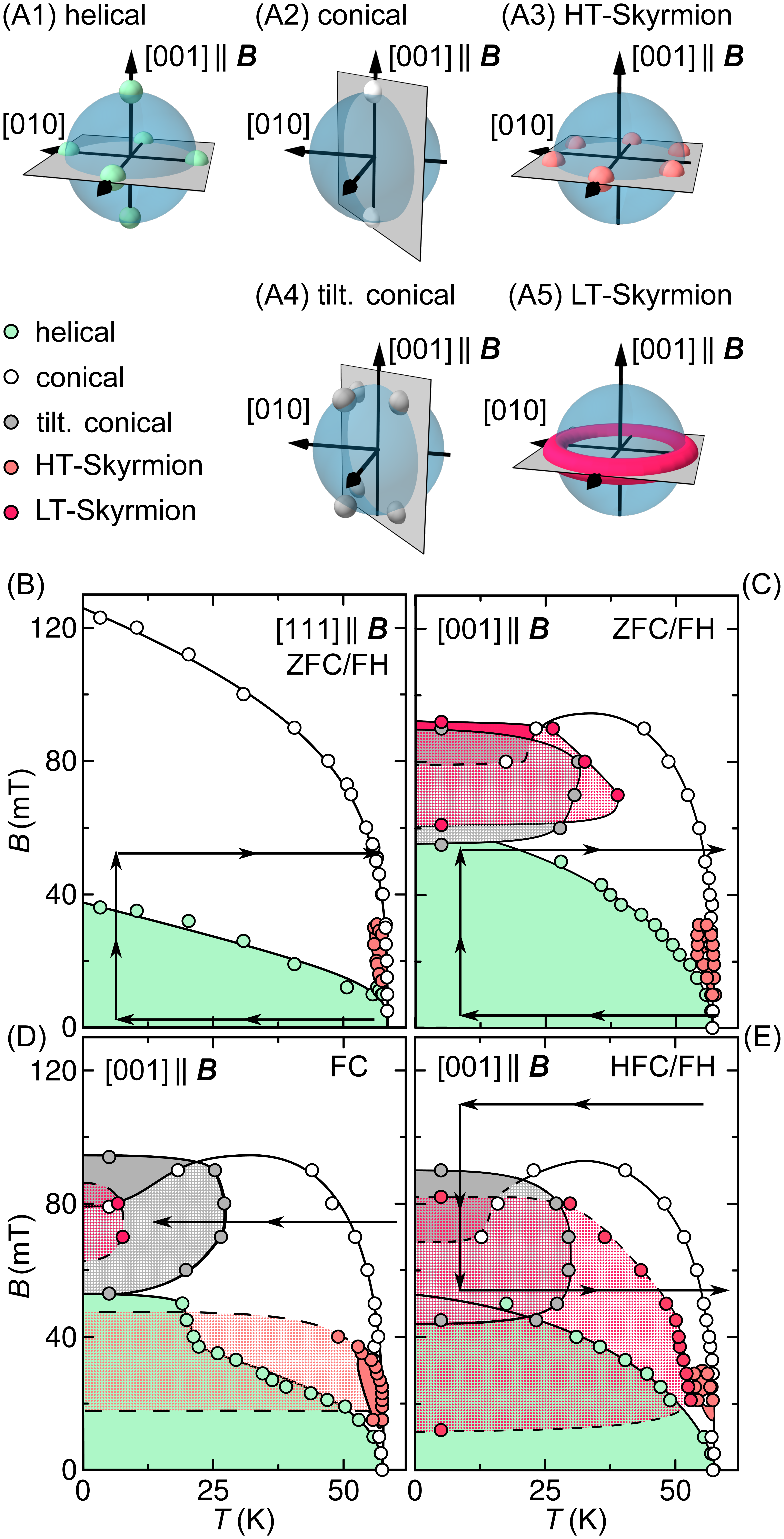}}
\bigskip
\bigskip

\Large\textbf{Fig.~1}

\clearpage

\begin{figure}[h]

\caption{
\textbf{Qualitative depiction of the intensity patterns in reciprocal space characterising the modulated magnetic order in {\cso} and magnetic phase diagrams for different temperature versus field histories.}  A: Intensity patterns of the helical order (green)(A1), conical order (white)(A2), high-temperature skyrmion phase (HT-Skyrmion, orange)(A3), tilted conical order (gray)(A4), and low-temperature skyrmion phase (LT-Skyrmion, red)(A5). The blue-shaded surface and the grey-shaded plane describe the manifold of $\vec{Q}$ vectors and the scattering plane, respectively. The scattering condition is satisfied at the intersection of both (note the field direction with respect to the scattering plane). (B) Magnetic phase diagram observed for ZFC/FH and field parallel to $\langle 111 \rangle$. (C)--(E): Magnetic phase diagrams observed for ZFC/FH, FC, and HFC/FH, respectively, for field parallel to $\langle 100 \rangle$. A highly hysteretic skyrmion phase emerges for all protocols at low temperatures (red shading), initiated by a pronounced tilting of the conical state (grey shading), where the hatched shading represents coexistence.} 
\label{figure1}
\end{figure}
\clearpage
\thispagestyle{empty}

%%%%%%%%%%%%%%%%%%%%%%%%%%%
\centerline{\includegraphics[width=0.6\textwidth,clip=]{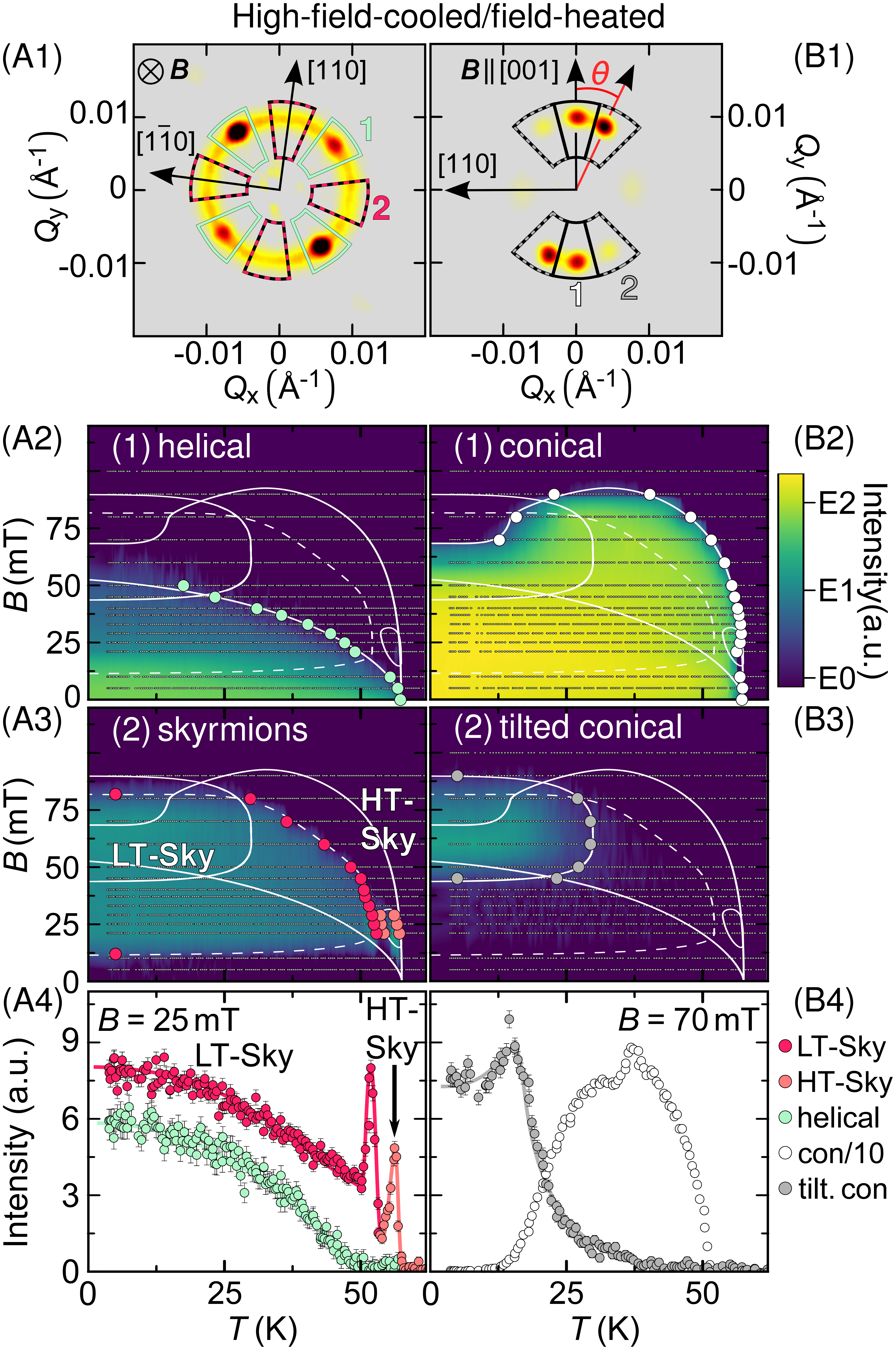}}
\bigskip
\bigskip

\Large\textbf{Fig.~2}
\clearpage

\begin{figure}[h]

\caption{\textbf{Typical small-angle neutron scattering patterns, intensity maps, and specific temperature dependences of the integrated intensity for HFC/FH.} We cool at $H>H_{c2}$ to base, reduce the field to the value of interest, and record data while heating (see Supplementary Information for data on other protocols). (A1), Typical intensity pattern for field parallel to the neutron beam. The intensities in sectors 1 and 2 correspond to the helical state, and the low-temperature (LT) and high-temperature (HT) skyrmion states, respectively. (B1), Typical intensity pattern for field perpendicular to the neutron beam. The intensities in sectors 1 and 2 correspond to the conical and the tilted conical states, respectively. (A2), (A3), (B2), and (B3), Intensity maps recorded for HFC/FH of the helical, the skyrmion, the conical, and tilted conical states, respectively. White lines mark the phase boundaries as shown in Fig.\,\ref{figure1}; tiny black dots mark locations where the data were recorded. (A4), Temperature dependence of the integrated intensity of the helical and the skyrmion states at $B=25\,{\rm mT}$. Note that the character of the skyrmion signal changes from LT to HT at the intensity minimum $\sim50\,{\rm K}$. (B4), Temperature dependence of the integrated intensity at $B=70\,{\rm mT}$ of the conical and the tilted conical states. All error bars represent plus and minus one standard deviation.
} \label{figure2}
\end{figure}
\clearpage
\thispagestyle{empty}

%%%%%%%%%%%%%%%%%%%%%%%%%%%
\centerline{\includegraphics[width=0.6\textwidth,clip=]{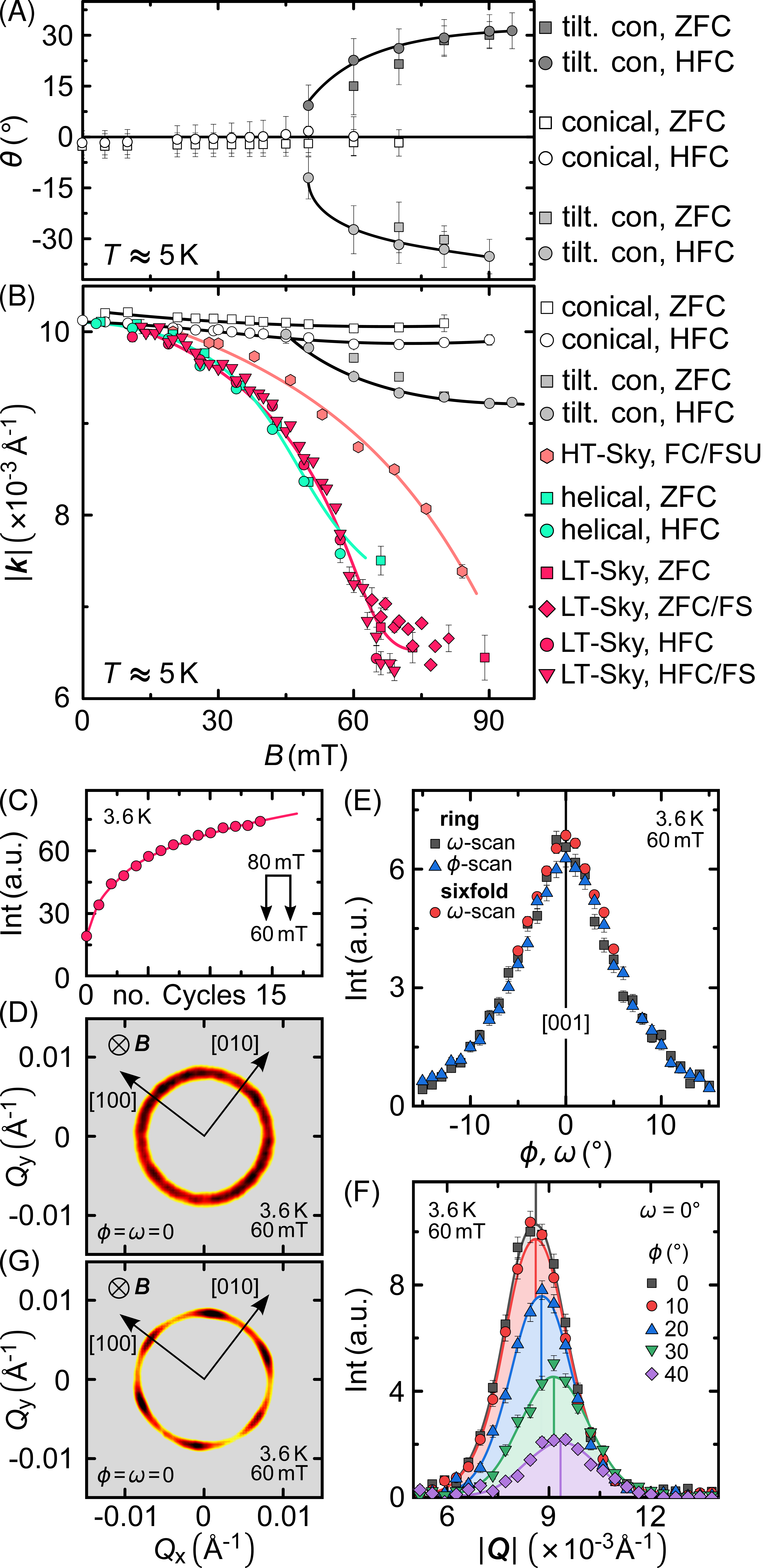}}
\bigskip
\bigskip

\Large\textbf{Fig.~3}
\clearpage

\begin{figure}[h]

\caption{
\textbf{Signatures of increasing magnetic anisotropies under increasing magnetic field at approximately $3.6\,{\rm K}$ and the skyrmionic nature of the low-temperature phase.} (A), Angle $\theta$ of the conical and the tilted conical modulation with respect to the field axis. 
(B), Modulus of the modulation $\vert \vec{k} \vert$ as a function of magnetic field for different phases and measurement protocols. For magnetic fields perpendicular to the modulations, $\vert \vec{k} \vert$ decreases, and four situations may be distinguished: conical state parallel to the field (white); $\vert \vec{k} \vert$ is unchanged; tilted conical state (grey), where as the tilting sets in $\vert \vec{k} \vert$ decreases; super-cooled high-temperature skyrmion state (orange), where $\vert \vec{k} \vert$ decreases; and helical state (green) and low-temperature skyrmion state (red), where the same strong decrease of  $\vert \vec{k} \vert$ is observed regardless of the measurement protocol. (C), Increase of intensity of the low-temperature skyrmion phase after high-field cooling as a function of the number of field cycles from 60\,mT to 80\,mT and back. (D), Typical ring of intensity of the low-temperature skyrmion phase at 60\,mT. (E), Rocking scans ($\omega$-scan) and rotation scan ($\phi$-scan) of the low-temperature skyrmion intensity with respect to an $\langle100\rangle$ axis. (F), Intensity distribution as a function of momentum dependence of the low-temperature skyrmion phase under various rotation angles $\omega$ at 60\,mT. (G), Typical six-fold intensity distribution of the low-temperature skyrmion phase following a rotation of the crystal by an angle $\phi>15^{\circ}$ away from $\langle100\rangle$.
All error bars represent plus and minus one standard deviation.
}
\label{figure3}
\end{figure}
\clearpage
\thispagestyle{empty}

%%%%%%%%%%%%%%%%%%%%%%%%%%%
\centerline{\includegraphics[width=0.6\textwidth,clip=]{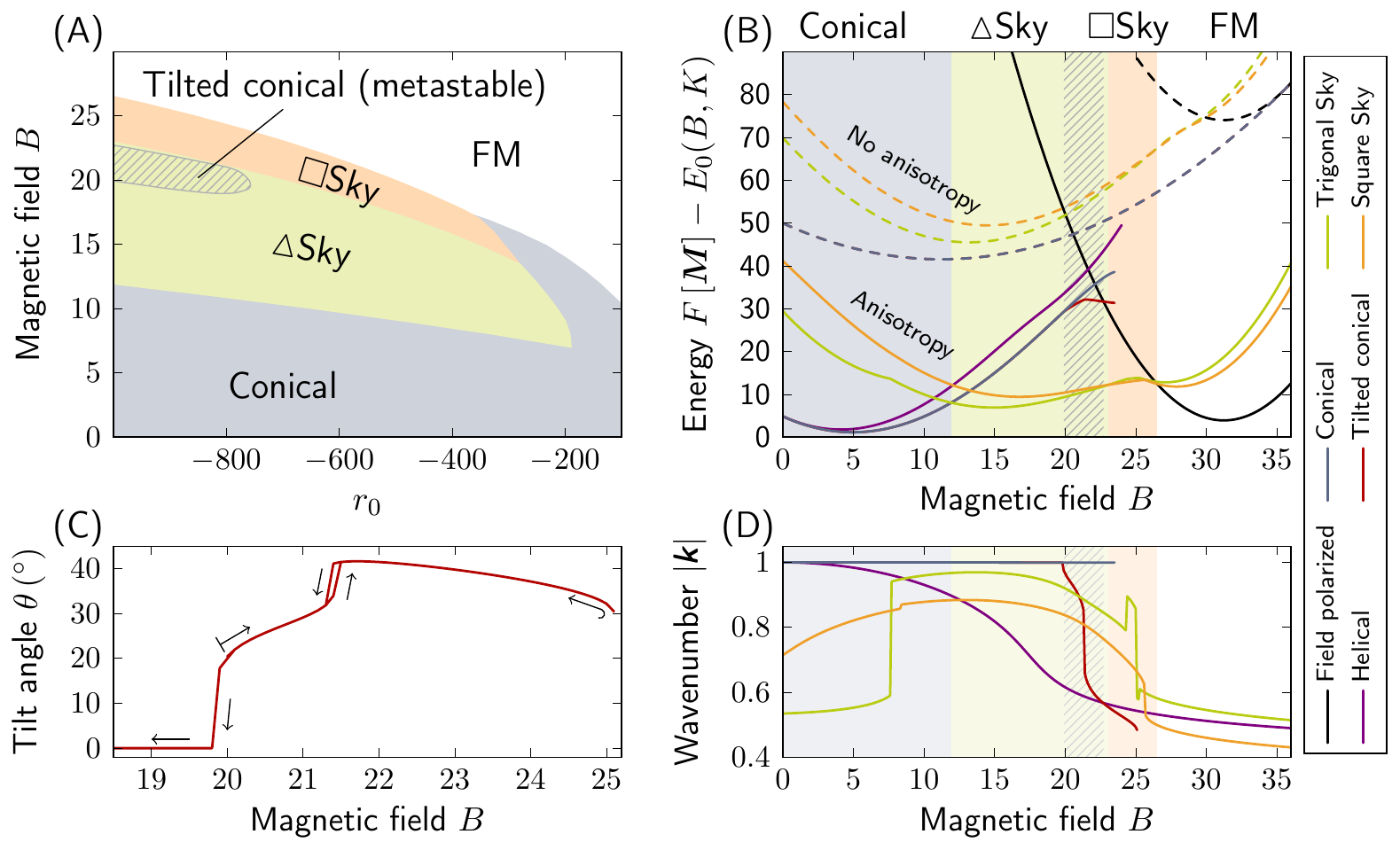}}
\bigskip
\bigskip

\Large\textbf{Fig.~4}
\clearpage 

\begin{figure}[h]

\caption{\textbf{Key results of the theoretical calculations.} Here $B$, $k$ and $r_0$ are in units of $U$, $D$ and $J$. (A), Mean-field magnetic phase diagram as a function of $r_0$ and $B$. The skyrmion phase and the metastable tilted phase (hatched region) emerge only for low temperatures (large negative values of $r_0$) and finite magnetic field $B$ (parameter: $K=0.0004U$). (B), Energy difference between the various modulated states and the field-polarized state as a function of magnetic field $B$ for $r_0=-1000D^2/J$. For better visibility, a reference energy $E_0(B,K)$ has been subtracted (quadratic in $B$, additional constant shift for dashed curves). In contrast to the case without anisotropy (dashed curves), where only a conical and a polarized phase occur, a trigonal and a square skyrmion phase form the ground state for intermediate magnetic fields. In a small range of fields (hatched region, $19.9<B<22.7$), the tilted conical state has a lower energy than both the conical and the polarized states, also defining the hatched region in (A). (C), Tilting angle of the conical state with respect to the field direction, corresponding to the experimental results presented in Fig.\,\ref{figure3}(A). (D) Field dependence of the modulus $\vert \vec{k} \vert$ of the various magnetic states, corresponding to Fig.\,\ref{figure3}(B).
} \label{figure4}
\end{figure}
%%%%%%%%%%%%%%%%%%%%%%%%%%%%%%%%%%%

\end{document}

\newcommand{\AR}[1]{{{\bf AR:} \em #1}}
\newcommand{\CP}[1]{{{\bf CP:} \em #1}}
\newcommand{\OLD}[1]{{\tiny #1}}

\newcommand{\mfs}{Mn$_{1-x}$Fe$_{x}$Si}
\newcommand{\mcs}{Mn$_{1-x}$Co$_{x}$Si}
\newcommand{\fcs}{Fe$_{1-x}$Co$_{x}$Si}
\newcommand{\cso}{Cu$_{2}$OSeO$_{3}$}

\newcommand{\rxx}{$\rho_{xx}$}
\newcommand{\rxy}{$\rho_{xy}$}
\newcommand{\rxytop}{$\rho_{\rm xy}^{\rm top}$}
\newcommand{\Drxyt}{$\Delta\rho_{\rm xy}^{\rm top}$}
\newcommand{\Sxy}{$\sigma_{xy}$}
\newcommand{\Sxya}{$\sigma_{xy}^A$}

\newcommand{\bco}{$B_{\rm c1}$}
\newcommand{\bct}{$B_{\rm c2}$}
\newcommand{\bao}{$B_{\rm A1}$}
\newcommand{\bat}{$B_{\rm A2}$}
\newcommand{\beff}{$B_{\rm eff}$}
\newcommand{\bmm}{$B_{\rm m}$}
\newcommand{\bs}{$B_{\rm S}$}
\newcommand{\btop}{$B_{\rm top}$}
\newcommand{\bnfl}{$B_{\rm NFL}$}

\newcommand{\tc}{$T_{\rm c}$}
\newcommand{\tmax}{$T_{\rm max}$}
\newcommand{\ts}{$T_{\rm S}$}

\newcommand{\pc}{$p_{\rm c}$}

\newcommand{\mb}{$\mu_0\,M/B$}
\newcommand{\dmdb}{$\mu_0\,\mathrm{d}M/\mathrm{d}B$}
\newcommand{\ddmddb}{$\mathrm{\mu_0\Delta}M/\mathrm{\Delta}B$}
\newcommand{\cac}{$\chi_{\rm ac}$}
\newcommand{\rechi}{${\rm Re}\,\chi_{\rm ac}$}
\newcommand{\imchi}{${\rm Im}\,\chi_{\rm ac}$}

\newcommand{\ozz}{$\langle100\rangle$}
\newcommand{\ooz}{$\langle110\rangle$}
\newcommand{\ooo}{$\langle111\rangle$}
\newcommand{\too}{$\langle211\rangle$}

\renewcommand{\vec}[1]{{\bf #1}}

\bibliographystyle{nature}

%%%%%%%%%%%%%%%%%%%%%%%%%%%%%%%%%%%

\title{Supporting online material:\\
Observation of Two Thermodynamically Disconnected Skyrmion Phases in {\cso}}

\author{A. Chacon}
\affiliation{Physik Department, Technische Universit\"at M\"unchen, D-85748 Garching, Germany}

\author{L. Heinen}
\affiliation{Institut f\"ur Theoretische Physik, Universit\"at zu K\"oln, D-50937 K\"oln, Germany}

\author{M. Halder}
\affiliation{Physik Department, Technische Universit\"at M\"unchen, D-85748 Garching, Germany}

\author{A. Bauer}
\affiliation{Physik Department, Technische Universit\"at M\"unchen, D-85748 Garching, Germany}

\author{W. Simeth}
\affiliation{Physik Department, Technische Universit\"at M\"unchen, D-85748 Garching, Germany}

\author{S. M\"uhlbauer}
\affiliation{Heinz Maier-Leibnitz Zentrum (MLZ), Technische Universit\"at M\"unchen, D-85748 Garching, Germany}

\author{H. Berger}
\affiliation{\'Ecole Polytechnique Federale de Lausanne, CH-1015 Lausanne, Switzerland}

\author{M. Garst}
\affiliation{Institut f\"ur Theoretische Physik, Technische Universit\"at Dresden, D-01062 Dresden, Germany}

\author{A. Rosch}
\affiliation{Institut f\"ur Theoretische Physik, Universit\"at zu K\"oln, D-50937 K\"oln, Germany}

\author{C. Pfleiderer}
\affiliation{Physik Department, Technische Universit\"at M\"unchen, D-85748 Garching, Germany}

\date{\today}

\begin{abstract}
We present details of the theoretical calculations and the experimental methods used in our studies of the magnetic phase diagram in {\cso}. We also present additional data illustrating the generic nature of our results.
\end{abstract}

\maketitle

%\newpage
%%%%%%%%%%%%%%%%%%%%%%%%%%%%%%%%%%%%%%%%%%%%%
\newpage

\section{Ginzburg Landau Analysis}

In the following the theoretical model used to describe our results will be presented in detail. Pedagogical introductions to this model may be found in various publications \cite{2009:Muhlbauer:Science,2013:Milde:Science,2013:Buhrandt:PRB,2015:Schwarze:NatMater}. In the following the emphasis will be put on specific aspects of the magnetic anisotropy terms and their implications for our experimental observations. 

\subsection{General Framework}

%\begin{equation}
%F\left[\vec{M}\right]=F_0\left[\vec{M}\right]+F_d\left[\vec{M}\right]+F_a\left[\vec{M}\right]
%\end{equation}
%\begin{multline}
%F_a\left[\vec{M}\right]=\sum_{\vec{k}}\Bigg(-K\sum_{\vec{k}_2,\vec{k}_3,\vec{k}_4} \left(
%M_{\vec{k}}^x M_{\vec{k}_2}^x M_{\vec{k}_3}^x M_{\vec{k}_4}^x +\dots\right)\delta_{\vec{k}+\vec{k}_2+\vec{k}_3+\vec{k}_4,0}+\\+ c_1 \left(k_x^4 +k_y^4
%+k_z^4 \right)\vec{M}_{\vec{k}}\cdot \vec{M}_{-\vec{k}} + c_2 \left(k_x^2 M_{\vec{k}}^x
%M_{-\vec{k}}^x+ \dots\right)\Bigg)
%\end{multline}

Our theoretical analysis is based on a Ginzburg-Landau $\phi^4$-Model. The free energy functional we use can be split into three parts $F=F_0+F_d+F_a$, as follows
\begin{multline}
F_0\left[\vec{M}\right]=\sum_{\vec{k}}\Bigg(\frac{J}{2}(\vec{k}\cdot\vec{k})( \vec{M}_{\vec{k}}\cdot
\vec{M}_{-\vec{k}}) + i D \vec{M}_{-\vec{k}} \cdot \left(\vec{k} \times \vec{M}_{\vec{k}}
\right) +r_0 \vec{M}_{\vec{k}}\cdot \vec{M}_{-\vec{k}}+\\+U\sum_{\vec{k}_2,\vec{k}_3,\vec{k}_4}
(\vec{M}_{\vec{k}}\cdot\vec{M}_{\vec{k}_2})(\vec{M}_{\vec{k}_3}\cdot\vec{M}_{\vec{k}_4})\delta_{\vec{k}+\vec{k}_2+\vec{k}_3+\vec{k}_4,0}\Bigg)-\vec{B}\cdot\vec{M}_0
\end{multline}
This functional has been shown to reproduce key properties of chiral magnets, such as the occurrence of the helical, conical \cite{1980:Bak:JPhysCSolidState, 1980:Nakanishi:SolidStateCommun} and high temperature skyrmion phase \cite{2009:Muhlbauer:Science,2013:Milde:Science, 2013:Buhrandt:PRB,2015:Schwarze:NatMater}. The parameters in this functional are given by the exchange strength $J$, the strength of the Dzyaloshinsky-Moriya interactions $D$, the magnetic field $B$, and the Ginzburg-Landau coefficients $r_0$ and $U$. 

\begin{equation}
F_d\left[\vec{M}\right]=\tau\left(\vec{M}_0 N \vec{M}_{0}+\sum_{\vec{k}}(\vec{k}\cdot \vec{M}_{\vec{k}})
(\vec{k}\cdot \vec{M}_{-\vec{k}})/\vec{k}\cdot\vec{k}\right),
\end{equation}
represents the effects of dipolar interactions, where $\tau$ is the relative strength of dipolar interactions and $N$ is the demagnetization tensor, with $\operatorname{tr}(N)=1$. 

\begin{equation}
F_a\left[\vec{M}\right]=-K\sum_{\vec{k},\vec{k}_2,\vec{k}_3,\vec{k}_4} \left(
M_{\vec{k}}^x M_{\vec{k}_2}^x M_{\vec{k}_3}^x M_{\vec{k}_4}^x +\dots\right)\delta_{\vec{k}+\vec{k}_2+\vec{k}_3+\vec{k}_4,0}
\end{equation}
represents the cubic anisotropy, where $K$ is the anisotropy constant. It turns out that other anisotropy terms, such as $\left(k_x^4+\dots\right)\vec{M}_{\vec{k}}\cdot \vec{M}_{-\vec{k}}$ and $k_x^2 M_{\vec{k}}^x M_{-\vec{k}}^x+ \dots$, are not needed for a qualitative understanding of our experimental results, as explained in the main text. The parameters $J$, $D$ and $U$ can be eliminated from the theory by a simple rescaling transformation. We therefore set $J=D=U=1$ in the following discussion. For \cso{} a value of $\tau\approx\chi^{\mathrm{int}}_{\mathrm{con}}/2\approx0.88$ has been reported before \cite{2015:Schwarze:NatMater}. This leaves $B$, $r_0$, $K$ and $N$ as the free parameters of our model, where  we use $r_0=-1000$, $K=0.0004$ and $N=\frac{1}{3}\mathds{1}$ unless stated otherwise.

\begin{figure}
	\begin{center}
		\includegraphics[width=0.9\textwidth]{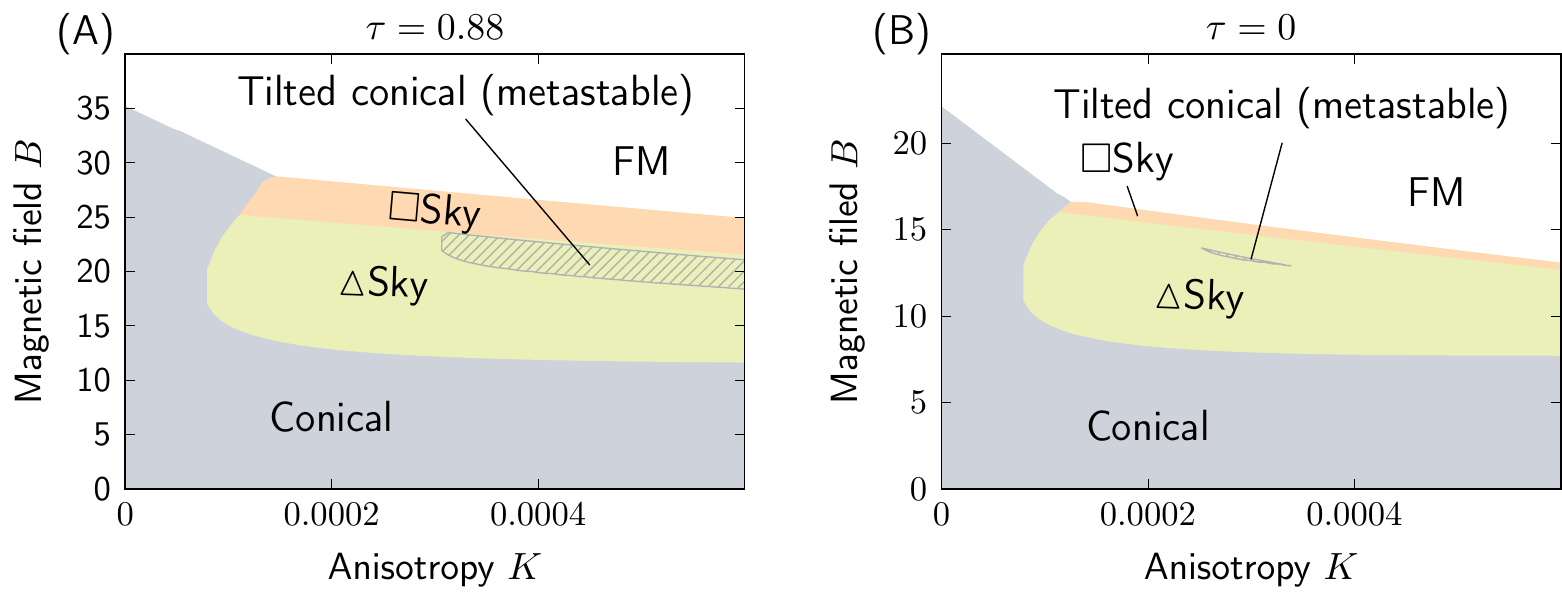}
	\end{center}
	\caption{Phase diagram obtained from Ginzburg-Landau theory as a function of anisotropy $K$ and magnetic field $B$ for two values of $\tau$. Beyond a critical strength of the anisotropy $K$, two different skyrmion lattices are stabilized for finite magnetic fields. Upon further increasing $K$, the energy of the tilted conical state drops below that of both conical and polarized states (hatched region). For $\tau=0.88$ this region is many times larger than for $\tau=0$. Parameters are $r_0=-1000$ and $N=\frac{1}{3}\mathds{1}$.
		\newline
		%\textit{figureSPhaseDiagramKB}
		\label{figureSPhaseDiagramKB}
	}
\end{figure}

As the focus of our study concerns the low-temperature limit, we neglect fluctuations and consider a mean-field approximation, i.e., we search for local minima of $F\left[\vec{M}\right]$. For this purpose we define the magnetization~$\vec{M}_{\vec{k}}$, parameterized such that it respects the relevant symmetries on a lattice in momentum-space corresponding to a specific phase. It should be noted that in general such a parametrization still allows for states beyond the phases of interest. For example in most cases the polarized state can be seen as a special case with infinite wavevector. Furthermore there are often several local minima close to each other in parameter space. This renders the identification of the global minimum highly nontrivial. We therefore use a combination of different methods to generate starting values for a quasi-Newton minimization. These include
\begin{itemize}
	\item random starting values
	\item manually scripted starting values
	\item the results of previous minimizations with similar external parameters
	\item interpolation of several previous results
\end{itemize}
A combination of these methods was used to establish and confirm the solution for each parametrization at each point in phase space. The resulting phase diagram as a function of $K$ and $B$ ($r_0$ and $B$) is shown in Fig.~\ref{figureSPhaseDiagramKB} (Fig.~4(A) of the main text).

%%%%%%%%%%%%%%%%%%%%%%%%%%%%%%%%%%%
\subsection{Tilted Conical Phase}

For the leading order anisotropy term discussed in the main text we find a skyrmion lattice ground state in a finite field range, accompanied by a metastable tilted conical phase, where the latter requires the additional effect of dipolar interactions or finely tuned parameters. In the following we present the energetics of this tilted conical state in further detail. 

\begin{figure}
	\begin{center}
		\includegraphics[width=0.5\textwidth]{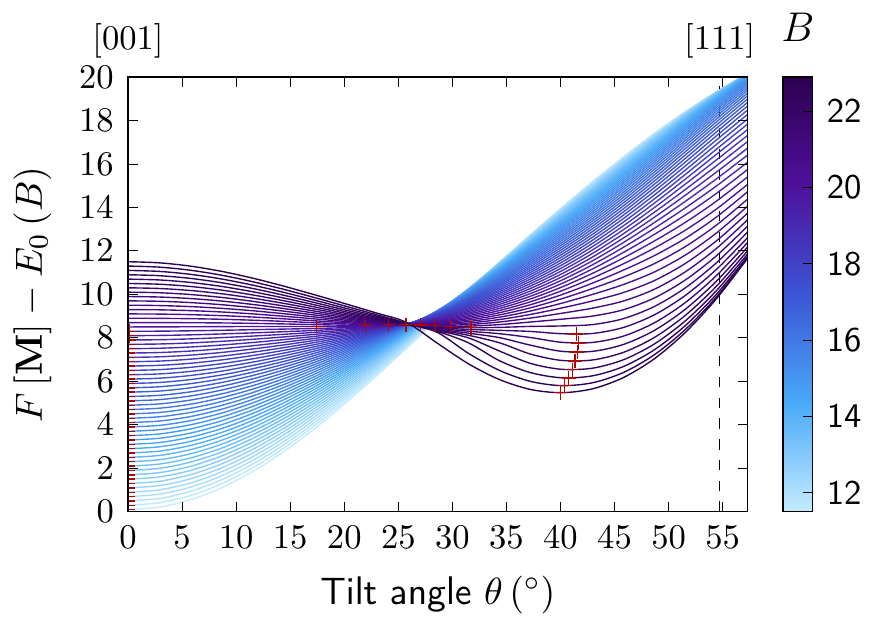}
	\end{center}
	\caption{Energy of the conical state as a function of the direction of $\vec{k}$ (the tilt angle $\theta$ away from $\left[001\right]$ towards $\left[111\right]$) for several values of the magnetic field $B$. For better visibility an arbitrary offset $E_0(B)$ was subtracted from each curve. Above $B_{ct}\approx19.9$ the minimum (red crosses) shifts towards finite angles $\theta\neq0$. Parameters are $r_0=-1000$, $K=0.004$, $\tau=0.88$ and $N=\frac{1}{3}\mathds{1}$. See Fig.\ref{figureSTiltAngleOfBAndK} (4(C) in main text) for the field dependence of tilt angle.
		\newline
		%\textit{figureSEnergyOfAngleAndB}
		\label{figureSEnergyOfAngleAndB}
	}
\end{figure}

Fig.~\ref{figureSEnergyOfAngleAndB} displays the calculated energy of the conical state as a function of the angle $\theta$ between $\vec{k}$ and $\vec{B}\parallel\left[001\right]$ for various field values ($\theta$ is measured on a great circle going from $\left[001\right]$ to $\left[111\right]$). Above a critical field $B_{ct}\approx19.9$ a minimum develops for $\theta\neq0$ (red crosses), where $B_{ct}$ is a function of $K$ and $\tau$.  Above $B_{ct}$ the conical state with $\vec{k}\parallel\vec{B}$ becomes unstable and the tilted conical state with $\vec{k}\nparallel\vec{B}$ becomes energetically favorable. Analytically, this can be understood as discussed in the main text. 

For very weak dipolar interactions (small $\tau$) the regime in which the tilted conical state becomes energetically favorable is in most cases masked by the onset of the polarized phase: $B_{c2}\leq B_{ct}$, as the polarized phase gains energy by the anisotropy $F_a$ for $K>0$ to the same extent as the tilted conical phase. However, as the polarized state is penalized by dipolar interactions, increasing $\tau$ shifts $B_{c2}$ above $B_{ct}$ and the tilted conical phase appears. This may be seen by comparing the two panels of Fig.~\ref{figureSPhaseDiagramKB}. For $\tau=0$ (panel (B)) the tilted phase appears only for a finely tuned set of parameters.
%, whereas for $\tau=0.88$ (panel (A)) it appears for a much larger set of Parameters
The value of $\tau\approx0.88$, reported for \cso{} (panel (A)), is large and generates a much larger area in phase space, where the tilted conical state has a lower energy than both the conical and the polarized states. This corresponds to the region\added{s} represented by a hatched area in Figs.~\ref{figureSPhaseDiagramKB} and 4 of the main text. The full region of metastability of the tilted phase is much larger, even for $\tau=0$.

We note that for the model as described above we did not find any set of parameters where the tilted conical phase becomes the ground state. It is masked by a skyrmion lattice, and only metastable. However, considering additional anisotropy terms we found several extensions of the model that exhibit a tilted phase as a ground state for certain parameters. It is important to note that all of these extended models show a helical phase with the $\left[111\right]$-direction as the easy axis for $B=0$ and some values of $r_0$. As this is not consistent with the experiment, we conclude that these additional anisotropies are not the source of a possible stabilization of the tilted conical state as we observed in \cso{}.

\begin{figure}
	\begin{center}
		\includegraphics[width=0.5\textwidth]{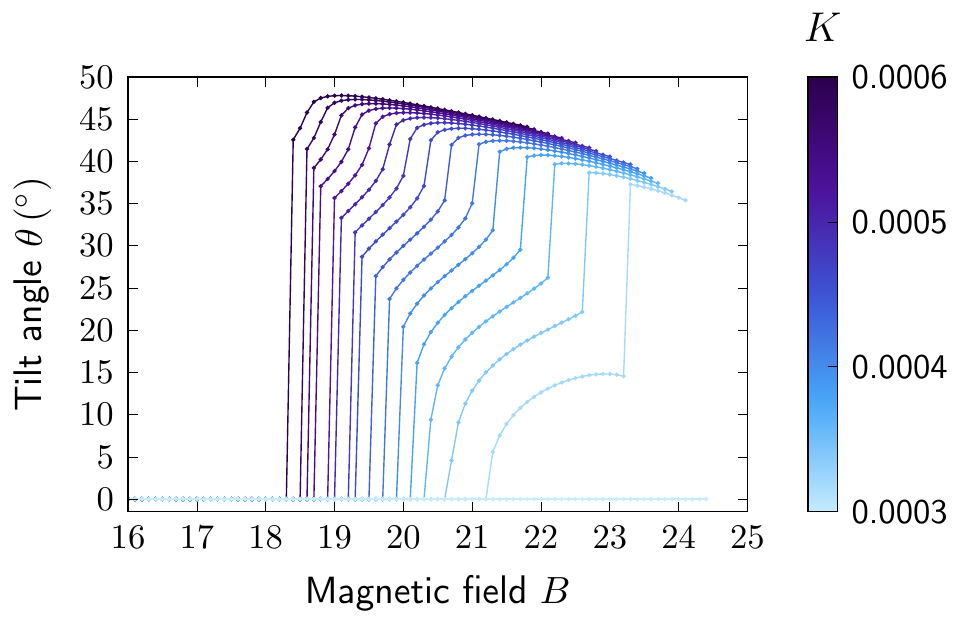}
	\end{center}
	\caption{Tilt angle $\theta$ of the tilted conical state as a function of the magnetic field $B$ for several values of $K$. Parameters are $r_0=-1000$, $\tau=0.88$ and $N=\frac{1}{3}\mathds{1}$.
		\newline
		%\textit{figureSTiltAngleOfBAndK}
		\label{figureSTiltAngleOfBAndK}
	}
\end{figure}

Further, from our model we calculated the tilt angle $\theta$ of the tilted conical phase as a function of $B$ and $K$. The results, shown in Fig.~\ref{figureSTiltAngleOfBAndK} for several values of $K$, qualitatively agree with our experimental observations. Namely, the angle increases over most of the field range of the tilted phase. However, the results in addition display one or two discontinuous jumps in tilt angle, depending on the value of $K$, which were not observed in our measurements.

\begin{figure}
	\begin{center}
		\includegraphics[width=0.5\textwidth]{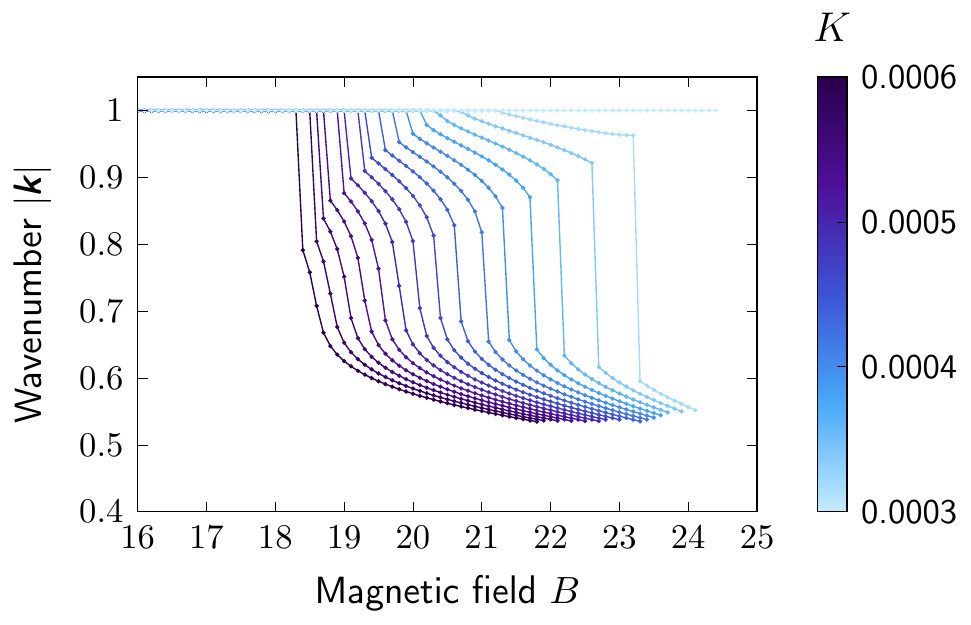}
	\end{center}
	\caption{Wavenumber $|\vec{k}|$ of the tilted conical state as a function of the magnetic field $B$ for several values of $K$. Parameters are $r_0=-1000$, $\tau=0.88$ and $N=\frac{1}{3}\mathds{1}$.
		\newline
		%\textit{figureSkConeOfBAndK}
		\label{figureSkConeOfBAndK}
	}
\end{figure}

These discontinuities not only show up in the tilt angle, but also in the wavenumber~$|\vec{k}|$ (see Fig.~\ref{figureSkConeOfBAndK}) and in the contribution of higher harmonics. They correspond to sudden changes in the magnetic microstructure of the tilted phase. As the details of these jumps and the values of the critical fields are very sensitive to the presence of further anisotropies, they may be smeared out by the presence of small amounts of disorder in our experiments.

%%%%%%%%%%%%%%%%%%%%%%%%%%%%%%%%%%%%%%%
\subsection{Low Temperature Skyrmion Lattice}

As discussed in the main text, a skyrmion lattice becomes energetically more favorable ($F_a$ gains energy for positive $K$) with increasing $K$. This ultimately leads to the stabilization of \emph{at least two} different skyrmion lattices if $K$ is large enough: a triangular and a square lattice (cf.\ Figs.~\ref{figureSPhaseDiagramKB} and 4 of the main text).
In our dimensionless units the threshold value for the cubic anisotropy in Fig. S1 above which the skyrmion phases appear corresponds approximately to $K_c \approx 0.0001$. 

Our considerations in terms of dimensionless units may alternatively be expressed in terms of dimensionfull units. Here the threshold for the low temperature skyrmion phase corresponds to a ratio 
\begin{equation}
K_c/(\mu_0 H^{\rm int}_{c2} M_s)\approx 0.07
\end{equation}
where $K_c$ is the threshold value for the cubic anisotropy in units of energy density, $H^{\rm int}_{c2}$ is the critical field separating the conical from the field-polarized phase and $M_s$ is the saturated magnetization. As this ratio scales like the square of spin-orbit coupling $K/(\mu_0 H^{\rm int}_{c2} M_s) \sim \lambda_{\rm SOC}^2$ the new skyrmion phases are only stable if $\lambda_{\rm SOC}$ is sufficiently strong. The strength of the dipolar interactions, however, does not play an important role for the low temperature skyrmion phase, in contrast to the tilted conical phase (compare Fig.~\ref{figureSPhaseDiagramKB} (A) and (B)).

\begin{figure}
	\begin{center}
		\includegraphics[width=0.5\textwidth]{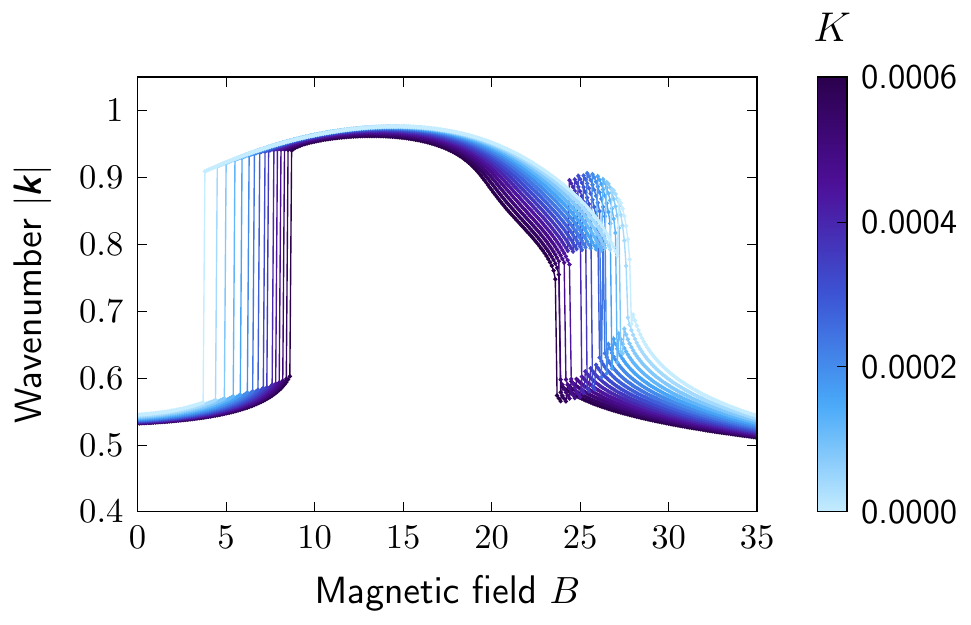}
	\end{center}
	\caption{Wavenumber $|\vec{k}|$ of the trigonal skyrmion lattice state as a function of the magnetic field $B$ for several values of $K$. For many parameters the lattice is characterized by a single $|\vec{k}|$. For others the lattice is deformed, so that a second wavevector is required for the description (omitted for clarity -- see Fig.~\ref{figureSkAndRealspaceSkx} for a partial plot). For example, the solutions with especially large $|\vec{k}|$ around $B=27$ are actually deformed almost into a square lattice. This is not surprising, since in this region the square lattice is energetically favorable. Parameters are $r_0=-1000$, $\tau=0.88$ and $N=\frac{1}{3}\mathds{1}$.
		\newline
		%\textit{figureSkSkxaOfBAndK}
		\label{figureSkSkxaOfBAndK}
	}
\end{figure}

\begin{figure}
	\begin{center}
		\includegraphics[width=0.5\textwidth]{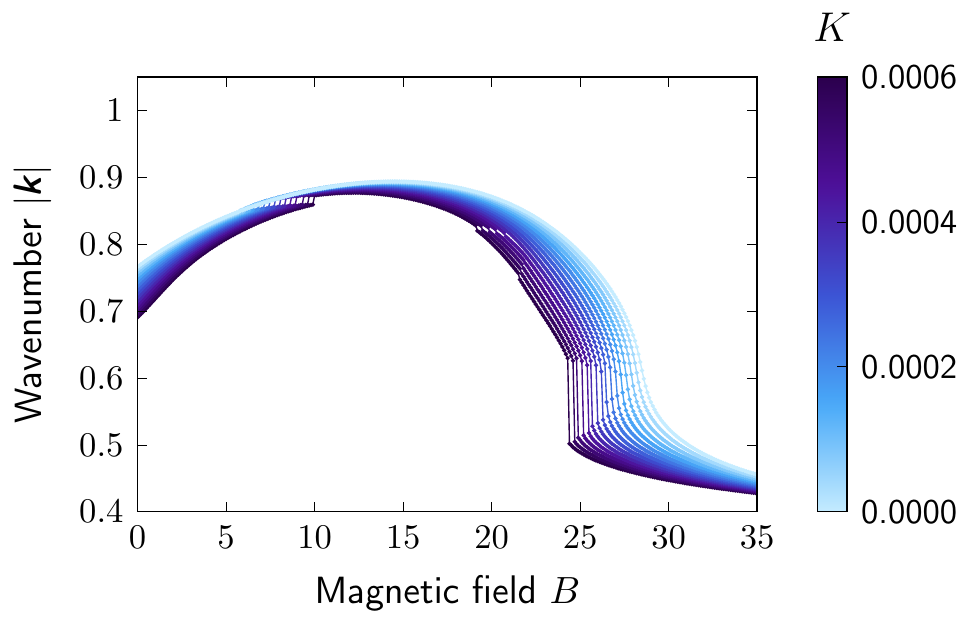}
	\end{center}
	\caption{Wavenumber $|\vec{k}|$ of the square skyrmion lattice state as a function of the magnetic field $B$ for several values of $K$. Around $K=0.0003$ a transition develops, separating a phase with smaller wavenumber from one with larger wavenumber. See also Fig.~\ref{figureSkAndRealspaceSkx}. Parameters are $r_0=-1000$, $\tau=0.88$ and $N=\frac{1}{3}\mathds{1}$.
		\newline
		%\textit{figureSkSkxsqrOfBAndK}
		\label{figureSkSkxsqrOfBAndK}
	}
\end{figure}

Similar to the tilted conical phase, both skyrmion lattice\added{s} develop as a function of magnetic field in a discontinuous manner.
%both the trigonal and the square skyrmion lattice morphologies emerge discontinuously as a function of magnetic field. 
This is shown with the help of the wavenumber $|\vec{k}|$ of the trigonal (square) lattice in Fig.~\ref{figureSkSkxaOfBAndK} (Fig.~\ref{figureSkSkxsqrOfBAndK}). Both figures show the corresponding wavenumber as a function of the external magnetic field $B$ for several values of $K$. For sufficiently strong anisotropy, both lattices exhibit more than one discontinuity in $|\vec{k}|$. In the case of the trigonal lattice, they correspond to lattice deformations towards either a square lattice or towards an elongation of the skyrmions. In the case of the square lattice they correspond mainly to different orientations of the lattice  either $\vec{k}\parallel\left[100\right]$ or $\vec{k}\parallel\left[110\right]$.

\begin{figure}
	\begin{center}
		\includegraphics[width=0.5\textwidth]{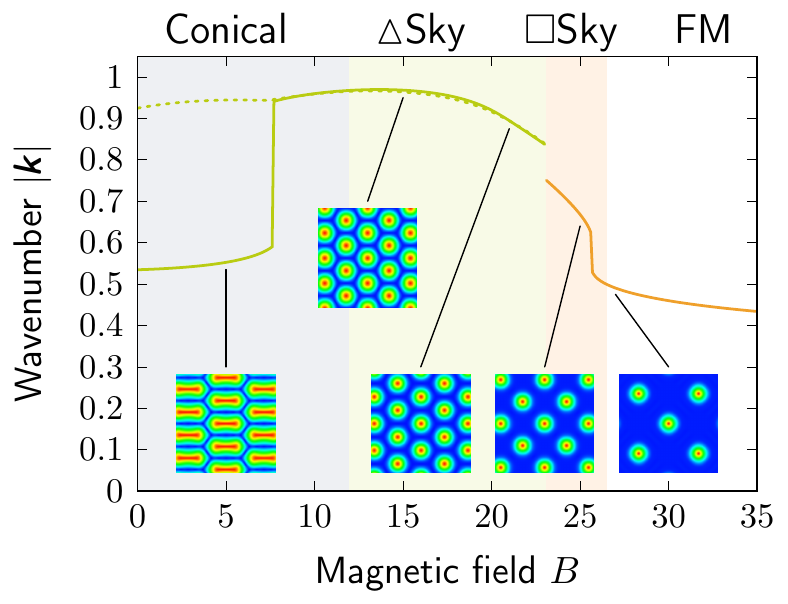}
	\end{center}
	\caption{Modulus of the wavenumber, $|\vec{k}|$, of the skyrmion lattice state with losest energy as a function of the magnetic field $B$. At $B\approx23$ a transition from a trigonal to a square lattice is observed. Below $B\approx7.7$ the trigonal lattice is strongly deformed by anisotropy, and characterized by two very different wavenumbers (solid/dashed lines). Background colors indicate the corresponding ground state. The insets show equal area real-space images for selected values of $B$. Parameters are $K=0.0004$, $r_0=-1000$, $\tau=0.88$ and $N=\frac{1}{3}\mathds{1}$.
		\newline
		%\textit{figureSkAndRealspaceSkx}
		\label{figureSkAndRealspaceSkx}
	}
\end{figure}

To illustrate the character of these different states, Fig.~\ref{figureSkAndRealspaceSkx} shows a series of real-space images of the skyrmion state with the smallest energy for different magnetic fields, together with the corresponding wavenumbers. The state changes from a strongly distorted trigonal skyrmion lattice for weak magnetic fields to an almost undistorted trigonal lattice, followed by two square lattices, with different skyrmion lattice constants. In the relevant region in phase space all of these states have similar energies, when compared to the other states in question.

We note that a square lattice of skyrmions has been reported before \cite{banerjee:PRX:14, 2015:lin:PRB, 2016:Utkan:PRB}. It was found in theoretical studies of magnetic single layers with easy-plane anisotropy. In bulk cubic chiral magnets like Cu$_2$OSeO$_3$ that are at the focus of our work such easy-plane anisotropies are however not present.
To our knowledge, only metastable square lattices of skyrmions have been previously reported in such systems \cite{2016:Karube:NatMater}.

%%%%%%%%%%%%%%%%%%%%%%%%%%%%%%%%%%%%
\clearpage
\section{Experimental Methods}

\subsection{Sample preparation}

High quality single-crystal {\cso} was grown by chemical vapor transport. Samples from the same batch were investigated in a large number of experimental studies. Where comparison is possible, all samples show consistently the same helimagnetic transition temperatures and characteristic field values taking into account sample shape and demagnetising fields. In particular, measurements of the magnetisation, ac susceptibility and specific heat of the specimen investigated in the SANS measurements reported here are in excellent agreement with the literature.

For the work reported in this paper a large single crystal was carefully polished into a sphere with a diameter of 2\,mm (Fig.\,\ref{figureS1}\,(a)). Excellent single crystallinity was confirmed at the neutron diffractometer HEIDI at FRM II. The sample was oriented using Laue x-ray diffraction and attached to the end of Al holder using GE varnish. Unfortunately a small angle ${\delta\sim \SI{8}{\degree}}$ between the vertical rotation axis and \hkl<110> was only noticed late during the experiments. It accounts for small differences of the integrated intensities observed in rocking scans, since the rocking axes were not perfectly parallel to the crystallographic axis. The small misalignment does not affect the results and conclusions reported in this paper.

\subsection{Small angle neutron scattering}

Small angle neutron scattering (SANS) measurements were performed at the beam line SANS-1 at FRM II \cite{muhlbauer:NIaMiPRSAASDaAE:16}. Neutrons with an incident wavelength  ${\lambda=\SI{7}{\AA}}$ were used with a FWHM wavelength spread of \SI{10}{\%}. The neutron beam was collimated over a distance of \SI{20}{\m} with a beam diameter of \SI{50}{\mm} at the entry of the collimation and a pinhole sample aperture with a diameter of 4\,mm located 350\,mm in front of the sample. The distance between sample and detector was \SI{20}{\m}. Taken together the resolution of this set for the azimuthal angle, modulus of the modulation and the rocking angle were ${\Delta \alpha=\SI{6.0}{\degree}}$, ${\Delta\vert \bm{Q}\vert=\SI{0.0011}{\AA^{-1}}}$, and ${\Delta \omega =\SI{0.14}{\degree}}$, respectively. The scattering pattern was recorded with an area-sensitive detector of \SI{1x1}{\m} equipped with 128 $^3$He tubes providing a spatial resolution of  \SI{8x8}{\mm} each.

\begin{figure}
\begin{center}
\includegraphics[width=0.7\textwidth]{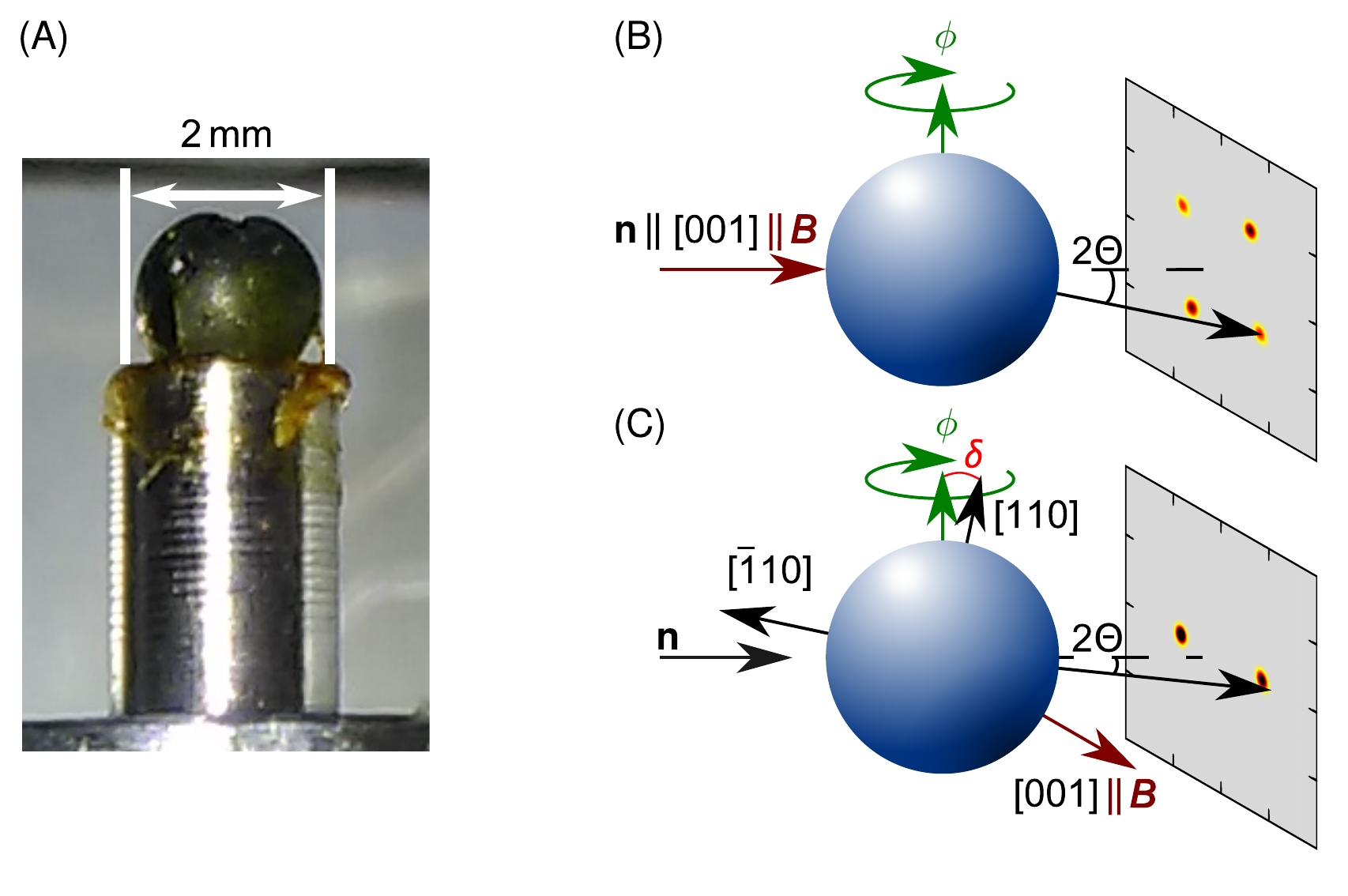}
\end{center}
\caption{Schematic depiction of the neutron scattering configurations used for our studies. (A) Spherical sample of \cso used for SANS measurements. (B) Set-up with magnetic field parallel to the incident neutron beam. This configuration permits to track modulations perpendicular to an applied magnetic field. It is typically used for studies of the helical state and the skyrmion state under magnetic field. (C) Set-up with magnetic field perpendicular to the incident neutron beam. This configuration permits to track modulations parallel to an applied magnetic field. The deviation of the crystallographic \hkl [110] from the axis of rotation is indicated by the angle $\delta$. See Fig.\,\ref{figureS1-2} for further information.
%\textit{figure-S1}
\label{figureS1}
}
\end{figure}

The sample as attached to the Al holder was mounted in a pulse-tube cooler (CCR) as combined with a cryogen-free \SI{5}{\tesla} superconducting magnet system, see Fig.~\ref{figureS1}. The sample temperature was measured with a Cernox sensor mounted in the immediate vicinity of the sample. Since data was recorded while sweeping the temperature continuously, the presence of small temperature gradients between the sample and the temperature sensor were determined in a set of systematic control measurements as a function of different sweep rates under cooling and heating. These measurements established, that the temperature gradients were vanishingly small at the lowest temperatures studied, and as high as a few \% around ${\sim\SI{60}{\kelvin}}$. Additional measurements were performed at temperatures down to 0.5\,K using a bespoke $^3$He system. These measurements served mostly to confirm the behaviour and phase diagrams down to 3.6\,K using the CCR.

Two configurations of the orientation of the magnetic field with respect to the neutron beam were used as shown in Fig.\,\ref{figureS1}\,(B) and (C), where the field was parallel and perpendicular to the incident neutron beam, respectively. The first configuration (Fig.\,\ref{figureS1}\,(B)) allowed to track scattering intensity perpendicular to the applied field. The second configuration (Fig.\,\ref{figureS1}\,(C)) allowed to track scattering intensity parallel to the field. The angle $\phi$ describes the angle between the [100] axis and the field direction.

\begin{figure}
\begin{center}
\includegraphics[width=0.5\textwidth]{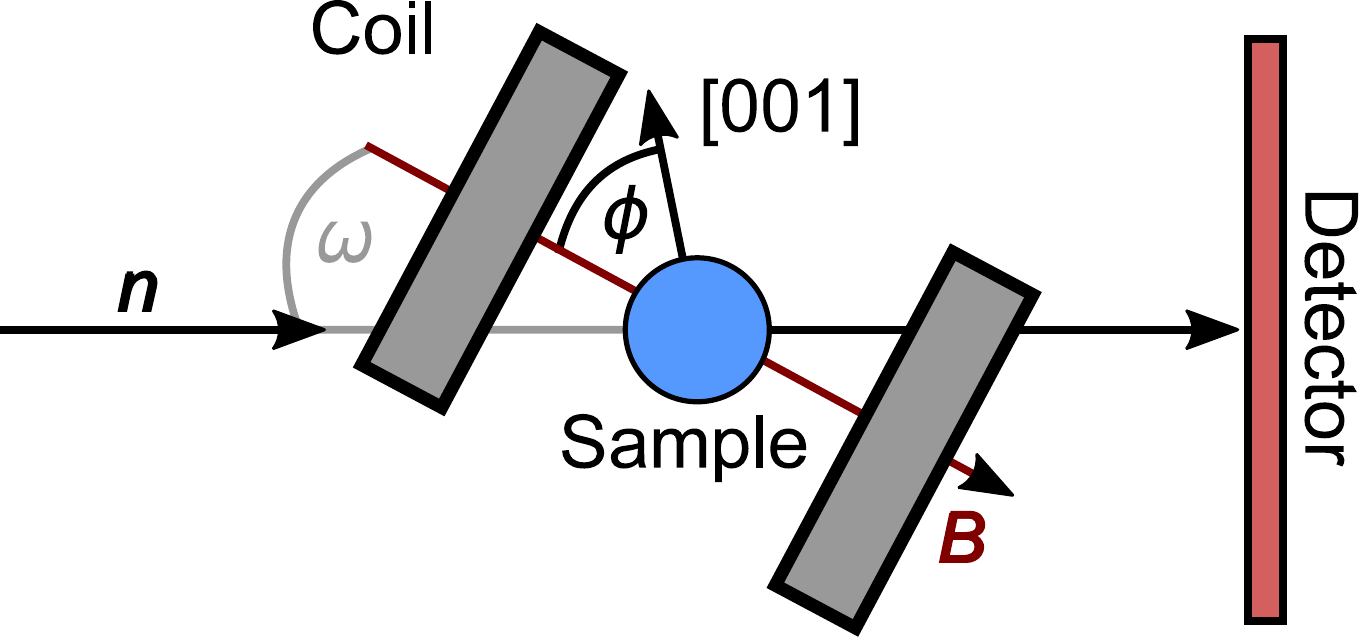}
\end{center}
\caption{
Schematic view as seen from the top of the neutron scattering configurations shown in Fig.\,\ref{figureS1}. 
The angle between the [100] axis and the field direction is denoted $\phi$, while $\omega$ represents the angle between magnetic field and neutron beam. The magnetic mosaicity may be determined by way of 'rocking scans', in which $\phi$ is kept constant while $\omega$ is varied. Thus the magnetic field direction in the sample is fixed throughout the rocking scan. 
\label{figureS1-2}
}
\end{figure}

Shown in Fig.\,\ref{figureS1-2} is a schematic view as seen from the top of the neutron scattering configuration. 
The angle $\omega$ defines the so called \textquoteleft rocking angle\textquoteright, by which both the sample and the applied magnetic field are rotated in order to measure the magnetic mosaicity. The angle $\phi$ describes the angle between the [100] axis and the field direction.  
For studies of the effects of a rotation of the crystallographic orientation against the applied field a bespoke sample stick was used, permitting accurate computer-controlled changes of the sample orientation with respect to the applied field and the neutron beam.

\begin{figure}[t]
\begin{center}
\includegraphics[width=1\textwidth]{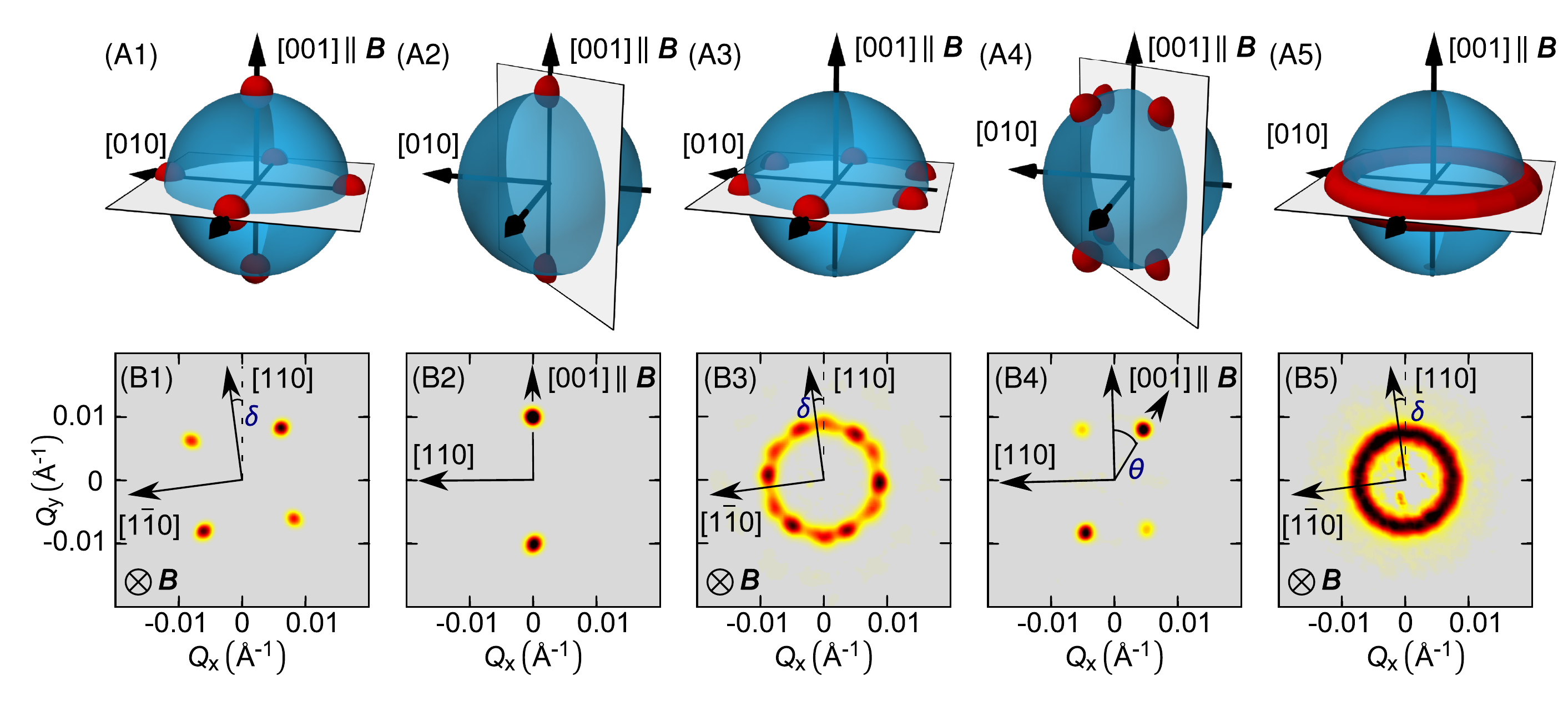}
\end{center}
\caption{Depiction of the intensity distributions characteristic of the different phases observed in our SANS studies. The following phases are distinguished: helical, conical, high-temperature skyrmion, tilted conical, and  low-temperature skyrmion. The scattering planes containing the defining key features of each phase are shown in gray shading. Note the magnetic field direction with respect to these scattering planes. Below each qualitative depiction shown are typical scattering patterns recorded in our measurements. Here $\delta$ defines the angle by which the rotation axis diverges from the \hkl [110] and $\theta$ the tilt angle of the tilted conical phase.
\newline
%\textit{figure-S2}
\label{figureS2}
}
\end{figure}

The scattering intensities observed in our studies may be fully accounted for in terms of five different contributions as illustrated in Fig.\,\ref{figureS2} (cf. Fig.\,1 in the main text). In these depictions the scattering plane containing key features is shown in gray shading (note the differences of field orientation with respect to these planes). The first scattering configuration (field parallel to the neutron beam, Fig.\,\ref{figureS1}\,(B)) is suitable to track details of the helical, high-temperature skyrmion and low-temperature skyrmion states, shown in Figs.\,\ref{figureS2}\,(A1), (A3), and (A5). The second scattering configuration (field perpendicular to the neutron beam, Fig.\,\ref{figureS1}\,(C)) is suitable to track details of the helical, conical and tilted conical states, shown in Figs.\,\ref{figureS2}\,(A1), (A2) and (A4). To track the combined changes of the scattering pattern as a function of temperature and magnetic field especially across phase transitions, the major part of our measurements were carried out for both scattering configurations accurately repeating the same temperature versus field histories.

Typical scattering patterns observed in our studies illustrating key features are shown in Figs.\,\ref{figureS2}\,(B1) through (B5). The diffraction pattern of the helical state, shown in Fig.\,\ref{figureS2}\,(B1), exhibits the four-fold symmetry with diffraction spots along the \hkl<100> axes. The pattern is rotated counter-clockwise with respect to the vertical direction representing the rotation axis due to the small misalignment of ${\delta\sim\SI{8}{\degree}}$ with respect to the \hkl <110> axis mentioned above. In the conical state the diffraction spots align accurately along the applied field (vertical direction) regardless of the small misalignment between the crystallographic axis and the field direction as shown in Fig.\,\ref{figureS2}\,(B2). A typical diffraction pattern of the high-temperature skyrmion phase is shown in Fig.\,\ref{figureS2}\,(B3). Twelve diffraction spots may be distinguished instead of the usual six diffraction spots. Consistent with the crystallographic orientation, these correspond to two domain populations of the skyrmion phase. 

Typical diffraction patterns of the two new characteristics observed in our studies are shown in Figs.\,\ref{figureS2}\,(B4) and (B5). First, as shown in Fig.\,\ref{figureS2}\,(B4) for sufficiently large magnetic fields along \hkl <100> the scattering peaks due to the conical state tilt away from the field direction (here the vertical axis) as denoted by the angle $\theta$. It is important to note that the small misalignment between the \hkl <110> axes and the rotation axis mentioned above, causes also a small misalignment of the field direction with respect to the \hkl <100> axis. In turn, this leads to a difference of the domain populations of the tilted conical states, which accounts for the differences of intensity of the spots shown in Fig.\,\ref{figureS2}\,(B4). Second, as shown in Fig.\,\ref{figureS2}\,(B5), for sufficiently large fields a ring of scattering intensity emerges perpendicular to the field direction. It is important to emphasize, that we do not observe any variation of the intensity as a function of azimuthal angle for this configuration. A sixfold pattern, associated with ordered skyrmion lattice domains, is only observed at lowest temperatures after careful sample preparation and a moderate rotation of $\phi$ exceeding $\sim15^{\circ}$ at 60\,mT. Empirically, this represents an important difference with the high-temperature skyrmion phase, which displays at least some azimuthal variation even for multi-domain configurations. 

\clearpage
%%%%%%%%%%%%%%%%%%%%%%%%%%%%%%%%%%%%%%%%%%%%%
\newpage

\subsection{Temperature versus field protocols}

All data were recorded following the temperature versus field protocols summarised in Fig.\,\ref{figureS3}. The only exception is the temperature versus field protocol described at the end of this supplement. Each measurement cycle started at a temperature of \SI{\sim70}{\kelvin} deep in the paramagnetic state above the helimagnetic transition at $T_c$. Prior to each measurement cycle keeping the sample at this high temperature, the superconducting magnet was carefully degaussed following the same procedure in order to minimise the amount of trapped flux. In the light of the strongly hysteretic effects displayed by the sample, this procedure proved to be important. As noted above the temperature of the sample was recorded with a calibrated Cernox sensor attached to the sample holder, where small gradients as high as a few \% observed at high temperatures were corrected. 

The following temperature versus field protocols were applied:
\begin{itemize}
\item \textbf{ZFC/FH}: The abbreviation refers to the expression zero-field-cooled/field-heated. The sample was cooled for zero magnetic field at an initial rate up to \SI{8}{\kelvin\minute^{-1}} down to the lowest temperature accessible \SI{\sim3.5}{\kelvin}. The magnetic field was swept to the field value of interest, $B_\textup{scan}$. Data was recorded at $B_\textup{scan}$ while continuously heating at a rate of \SI{2}{\kelvin\min^{-1}} up to \SI{65}{\kelvin}. During the temperature sweep data was recorded continuously  for periods of \SI{5}{\second} and stored, while the sweep continued. Storing the data caused a dead time of \SI{1}{\second}. Thus, each data point represents an average over a change of temperature of \SI{\sim83}{\milli\kelvin}. 

\item \textbf{FC}:
The abbreviation refers to the expression field-cooled. At high temperature before starting the scan the magnetic field was swept to the field value of interest, $B_\textup{scan}$. Data was recorded at $B_\textup{scan}$ while continuously cooling down to the lowest temperature accessible of \SI{\sim3.5}{\kelvin}. The initial cooling rate was as high as \SI{8}{\kelvin\minute^{-1}}. During the temperature sweep data was recorded continuously  for periods of \SI{5}{\second} and stored, while the sweep continued. Storing the data caused a dead time of \SI{1}{\second}. Thus, each data point represents an average over a change of temperature up to \SI{\sim333}{\milli\kelvin}. 

\item \textbf{HFC/FH}:\\
The abbreviation refers to the expression high-field-cooled/field-heated. At high temperature before starting the scan the magnetic field was swept to a value $B=\SI{250}{\milli\tesla}$, significantly higher than the upper critical field $B_{c2}$. The sample was cooled at an initial rate of up to \SI{8}{\kelvin\minute^{-1}} at $B=\SI{250}{\milli\tesla}$ down to the lowest temperature accessible of \SI{\sim3.5}{\kelvin}. The magnetic field was decreased to the value of interest, $B_\textup{scan}$. Data was recorded at $B_\textup{scan}$ while continuously heating at a rate of \SI{2}{\kelvin\min^{-1}} up to \SI{65}{\kelvin}. During the temperature sweep data was recorded continuously  for periods of \SI{5}{\second} and stored, while the sweep continued. Storing the data caused a dead time of \SI{1}{\second}. Thus, each data point represents an average over a change of temperature of \SI{\sim83}{\milli\kelvin}.

\begin{figure}[t]
\begin{center}
\includegraphics[width=0.85\textwidth]{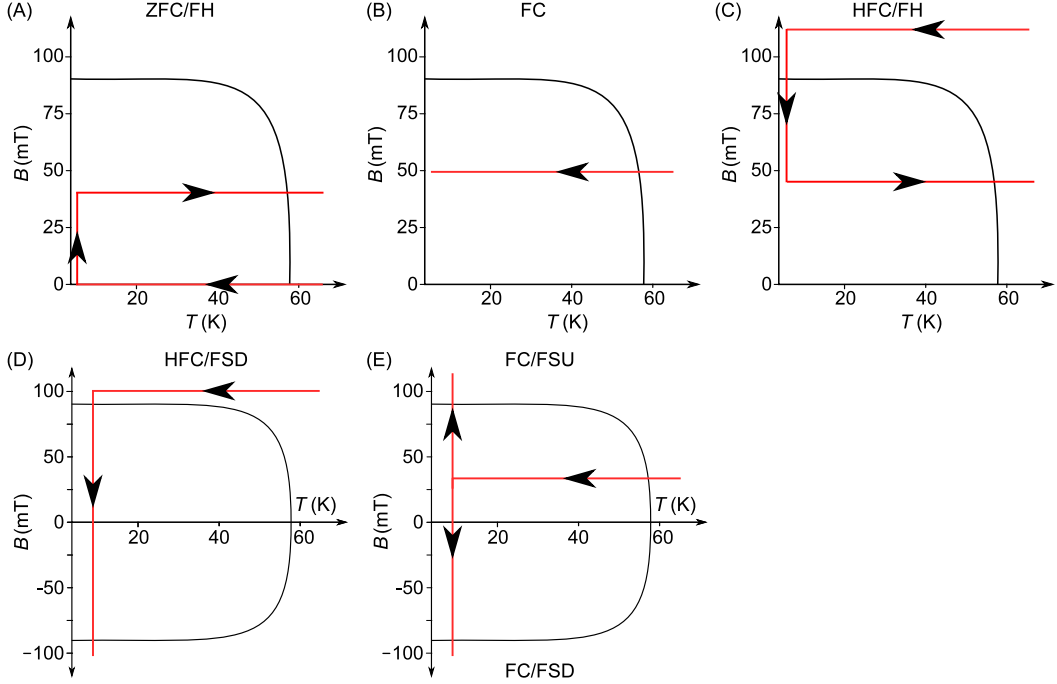}
\end{center}
\caption{Depiction of the temperature versus field protocols investigated in this study.  (A) ZFC/FH: zero-field-cooled/field-heated; (B) FC: field-cooled; (C) HFC/FH: high-field-cooled/field-heated: (D) HFC/FSD: high-field-cooled/field-sweep-down; (E) FC/FSU: field-cooled/field-sweep-up; FC/FSD: field-cooled/field-sweep-down.
\newline
%\textit{figure-S3}
\label{figureS3}
}
\end{figure}

\item \textbf{HFC/FSD}:
The abbreviation refers to the expression high-field-cooled/field-sweep-down. At high temperature before starting the scan the magnetic field was swept to value of $B=\SI{250}{\milli\tesla}$, significantly higher than the upper critical field $B_{c2}$. The sample was cooled at an initial rate of up to \SI{8}{\kelvin\minute^{-1}} at $B=\SI{250}{\milli\tesla}$ down to the  temperature of interest, $T_{scan}$. Data was recorded at $T_\textup{scan}$ while continuously decreasing the magnetic field at a rate of \SI{0.25}{\milli\tesla \second^{-1}} to a negative field below $-B_{c2}$. During the field sweep data was recorded continuously for periods of \SI{5}{\second} and stored, while the sweep continued. Storing the data caused a dead time of \SI{1}{\second}. Thus, each data point represents an average over a change of field of \SI{\sim1.25}{\milli\tesla}.

\item \textbf{FC/FSU}: 
The abbreviation refers to the expression high-field-cooled/field-sweep-up. 
At high temperature before starting the scan the magnetic field was swept to ${B_\textup{FC}=\SI{29}{\milli\tesla}}$. The sample was cooled at an initial rate of up to \SI{8}{\kelvin\minute^{-1}} and $B_\textup{FC}$ down to the  temperature of interest, $T_{scan}$. Data was recorded at $T_\textup{scan}$ while continuously increasing the magnetic field at a rate of \SI{0.25}{\milli\tesla \second^{-1}} to a positive field above $B_{c2}$. During the field sweep data was recorded continuously for periods of \SI{5}{\second} and stored, while the sweep continued. Storing the data caused a dead time of \SI{1}{\second}. Thus, each data point represents an average over a change of field of \SI{\sim1.25}{\milli\tesla}.

\item \textbf{FC/FSD}: 
The abbreviation refers to the expression field-cooled/field-sweep-down. 
At high temperature before starting the scan the magnetic field was swept to ${B_\textup{FC}=\SI{29}{\milli\tesla}}$. The sample was cooled at an initial rate of up to \SI{8}{\kelvin\minute^{-1}} at $B_\textup{FC}$ down to the  temperature of interest, $T_{scan}$. Data was recorded at $T_\textup{scan}$ while continuously decreasing the magnetic field at a rate of \SI{0.25}{\milli\tesla \second^{-1}} to a negative field below $-B_{c2}$. During the field sweep data was recorded continuously for periods of \SI{5}{\second} and stored, while the sweep continued. Storing the data caused a dead time of \SI{1}{\second}. Thus, each data point represents an average over a change of field of \SI{\sim1.25}{\milli\tesla}.

\end{itemize}

\clearpage
%%%%%%%%%%%%%%%%%%%%%%%%%%%%%%%%%%%%%%%%%%%%%
\newpage

\section{Further experimental results}
\subsection{Intensity maps based on temperature sweeps}

The magnetic phase diagrams presented in Fig.\,1 of the main text are based on a dense mesh of data recorded as a function of temperature and magnetic field following the temperature versus field protocols described above. Phase boundaries were either defined at the point where the intensity differed from the background by more than five standard deviations, $5\,\sigma$, or at clear changes of the diffraction patterns, see Fig.\ref{figureS2}.  Typical data recorded for HFC/FH have been presented in Fig.\,2 of the main text.

\begin{figure}[t]
\begin{center}
\includegraphics[width=0.5\textwidth]{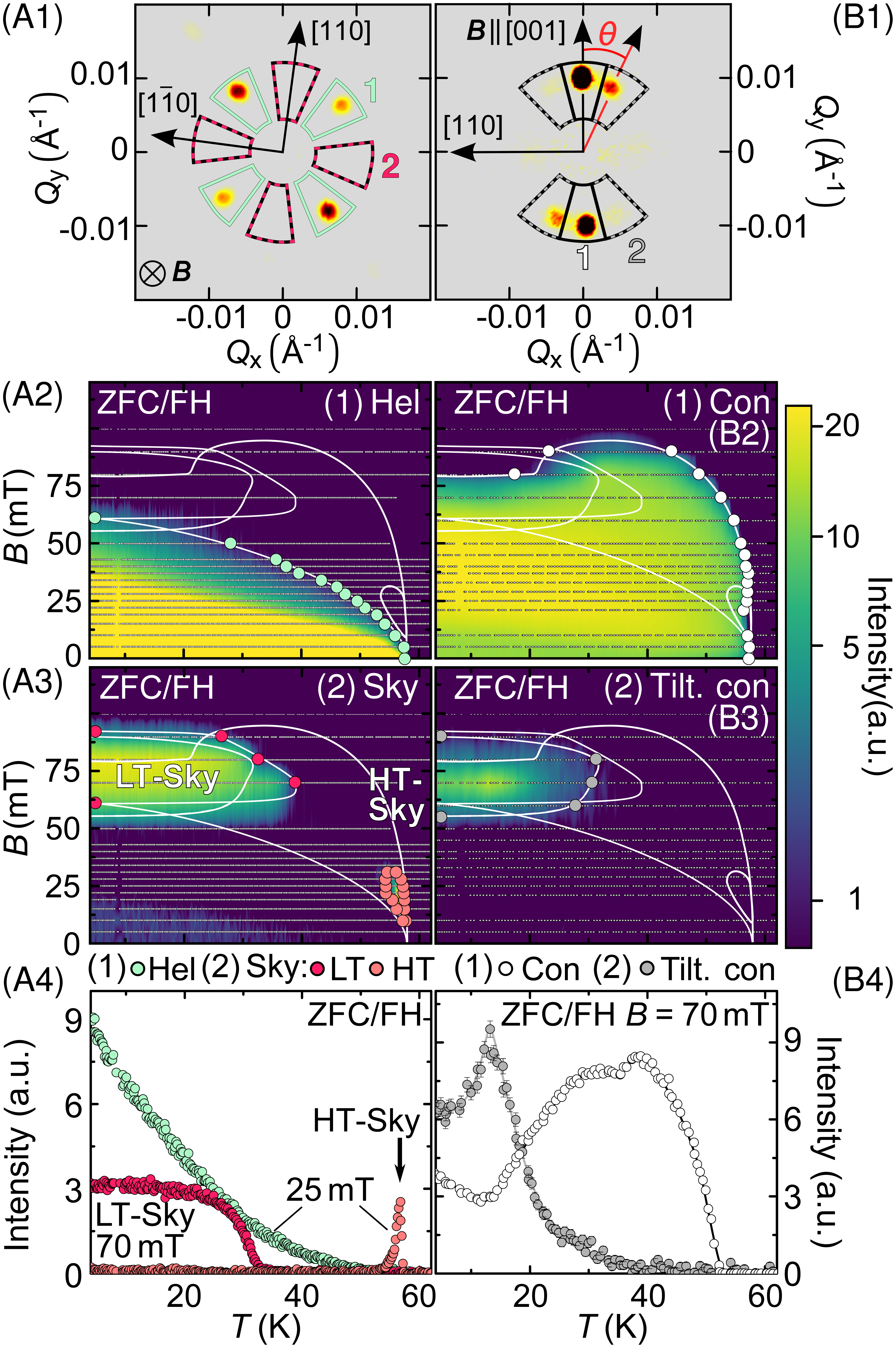}
\end{center}
\caption{{Typical SANS patterns, intensity maps, and specific temperature dependences of the integrated intensity for ZFC/FH  (A1) Typical intensity pattern for field parallel to the neutron beam. The intensities in sectors 1 and 2 correspond to the helical state and the low-temperature (LT) and high-temperature (HT) skyrmion states. Overlapping skyrmion state signal is corrected by substracting sector 2 from sector 1. (B1) Typical intensity pattern for field perpendicular to the neutron beam. The intensities in sectors 1 and 2 correspond to the conical and the tilted conical states, respectively. Panels (A2), (A3), (B2) and (B3): Intensity maps recorded for ZFC/FH of the helical, the skyrmion, the conical and tilted conical states, respectively. White lines mark the phase boundaries as shown in Fig.~1 of the main text; black dots mark locations were data was recorded. (A4) Temperature dependence of the integrated intensity of the helical and the high-temperature skyrmion state at $B=25\,{\rm mT}$, the data for the low-temperature were obtained at $B=\SI{70}{\milli\tesla}$. (B4) Temperature dependence of the integrated intensity at $B=70\,{\rm mT}$ of the conical and  tilted conical states.
}
\newline
%\textit{figure-S4a}
\label{figureS4a}
}
\end{figure}

\begin{figure}[t]
\begin{center}
\includegraphics[width=0.5\textwidth]{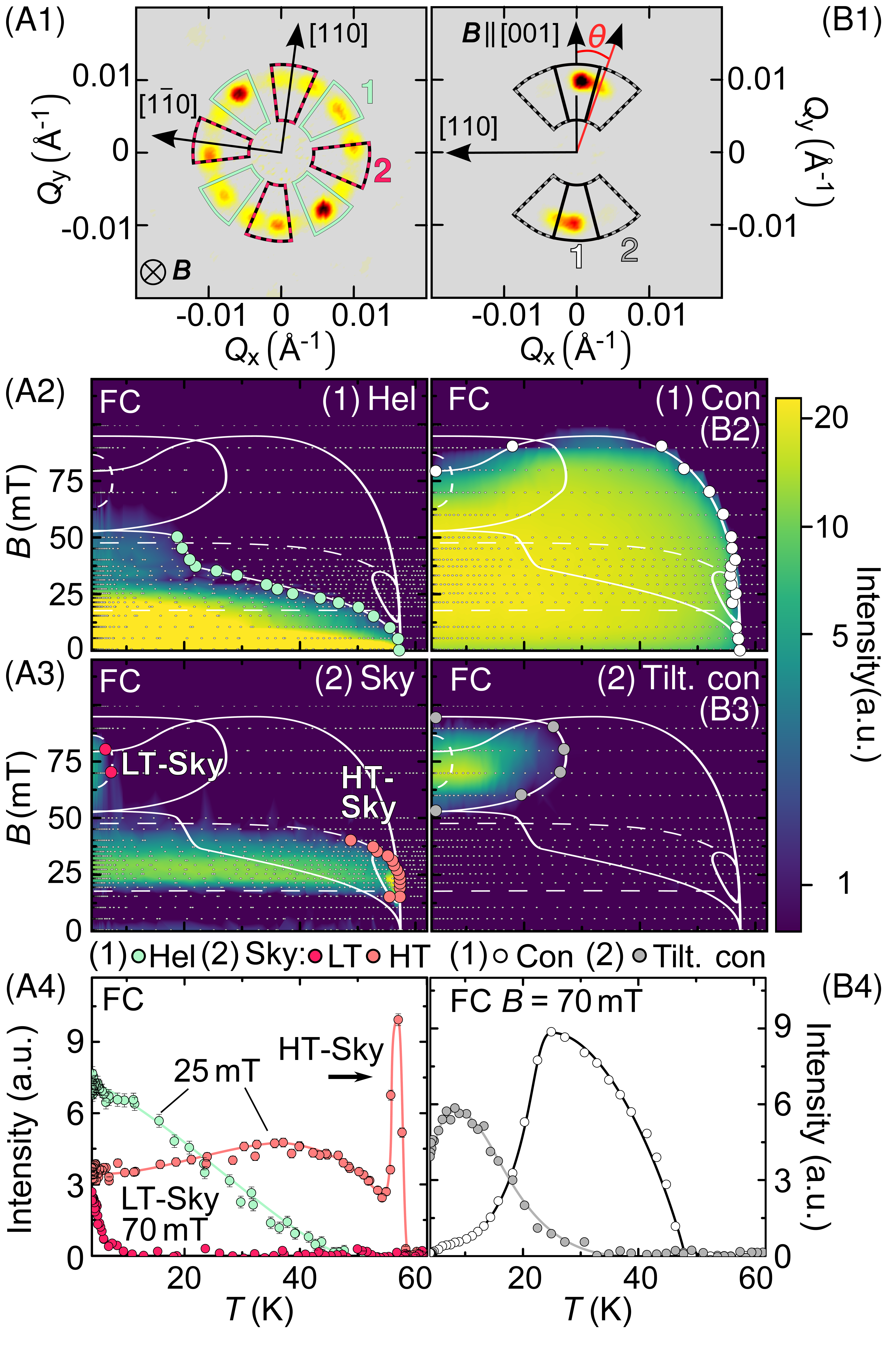}
\end{center}
\caption{
Typical SANS patterns, intensity maps, and specific temperature dependences of the integrated intensity for FC  (A1) Typical intensity pattern for field parallel to the neutron beam. The intensities in sectors 1 and 2 correspond to the helical state and the low-temperature (LT) and high-temperature (HT) skyrmion states. Overlapping skyrmion state signal is corrected by substracting sector 2 from sector 1. (B1) Typical intensity pattern for field perpendicular to the neutron beam. The intensities in sectors 1 and 2 correspond to the conical and the tilted conical states, respectively. Panels (A2), (A3), (B2) and (B3): Intensity maps recorded for ZFC/FH of the helical, the skyrmion, the conical and tilted conical states, respectively. White lines mark the phase boundaries as shown in Fig.~1 of the main text; black dots mark locations were data was recorded. (A4) Temperature dependence of the integrated intensity of the helical and the high-temperature skyrmion state at $B=25\,{\rm mT}$, the data for the low-temperature were obtained at $B=\SI{70}{\milli\tesla}$. (B4) Temperature dependence of the integrated intensity at $B=70\,{\rm mT}$ of the conical and  tilted conical states.
\newline
%\textit{figure-S4b}
\label{figureS4b}
}
\end{figure}

Intensity maps and selected temperature sweeps of the data recorded for ZFC/FH and FC are shown in Figs.\,\ref{figureS4a} and \ref{figureS4b}, respectively. The figures are organised in analogy with Fig.\,2 of the main text. Panels on the left hand side were recorded with the neutron beam parallel to the incident neutron beam. Panels on the right hand side were recorded with the neutron beam perpendicular to the incident neutron beam. Accordingly the intensity maps on the left hand side display the characteristics of the low-temperature and high-temperature skyrmion phase, as well as the helical state,  whereas the intensity maps on the right hand side display the characteristics of the conical and tilted conical state. These were obtained by integrating the intensity for each patterned measured over the sectors 1 and 2 shown on the first row. In case of coexistence between the helical phase and one of the skyrmion lattice, the former intensity was obtained by substracting the intensity of sector 2 from sector 1.

\newpage

%%%%%%%%%%%%%%%%%%%%%%%%%%%%%%%%%%%%%%%%%%%%%
%\newpage
%\section{Importance of temperature and field history}

\newpage

\clearpage
%%%%%%%%%%%%%%%%%%%%%%%%%%%%%%%%%%%%%%%%%%%%%

\subsection{Magnetic phase diagrams based on field sweeps}

\begin{figure}[t]
\begin{center}
\includegraphics[width=0.7\textwidth]{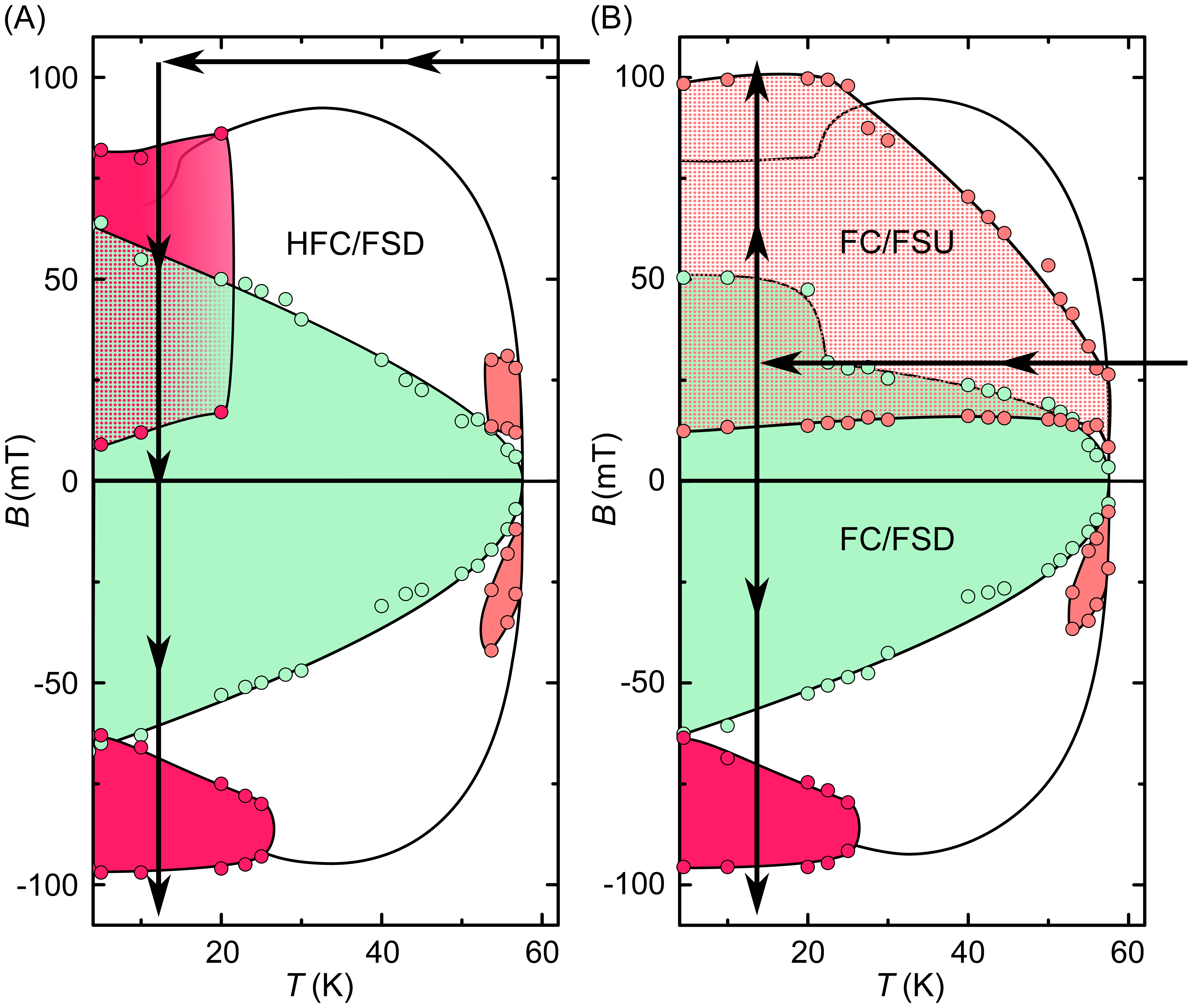}
\end{center}
\caption{
Magnetic phase diagrams inferred from field sweeps. Note that data was only recorded for field parallel to the neutron beam, i.e., no information is available on the conical and tilted conical states. Phase boundaries of the conical state are taken from the temperature sweeps (ZFC/FH, FC and HFC/FH), which establish reversible behaviour. (A) Magnetic phase diagram observed for the HFC/FSD protocol. For positive and negative field values the helical state displays essentially the same phase boundaries. In contrast, the low-temperature skyrmion phase is highly hysteretic, stabilising over a larger field range when approached from the field-polarised state as compared to stabilisation from the conical state. The high-temperature skyrmion phase shows small hysteresis. (B) Magnetic phase diagram observed for combined FC/FSU and FC/FSD measurements, where the field-cooling was carried out at 29\,mT in order to generate a super-cooled high-temperature skyrmion phase. This skyrmion phase survives only in the positive field region and transitions into a helical field around $\sim\SI{10}{\milli\tesla}$ at the lowest temperatures. The helical phase boundaries for positive fields resemble strongly those obtained from field-cooling (see main text), while it is not possible to distinguish between low- and high-temperature skyrmion phase for positive fields due to the strong field dependence of the modulation length $|\bm{k}|$. All phase boundaries for negative field values are essentially the same as those observed after the HFC/FSD protocol for negative fields (lower part of (A)), thus the magnetic state resets when the magnetic field orientation is reversed.
\newline
%\textit{figure-S4c}
\label{figureS4c}
}
\end{figure}

The magnetic phase diagrams for $B\parallel\hkl[100]$ inferred from field sweeps recorded as part of different temperature versus field protocols are shown in Fig.\,\ref{figureS4c}. For lack of beam time data was only recorded for field parallel to the incident neutron beam, i.e., no data were recorded of the conical and tilted conical states. For sake of comparison the phase boundaries of the conical with the field-polarised state, determined in the temperature sweeps reported in the main text, have been added to the phase diagram. It is helpful to note that the phase boundary between the conical and the field-polarised state observed in the various temperature sweeps is highly reversible.

Shown in Fig.\,\ref{figureS4c}\,(A) is the phase diagram observed under HFC/FSD. For positive and negative field values the helical and the high-temperature skyrmion phase display essentially the same phase boundaries. A key observation concerns the recovery of the helical state for small fields under decreasing field. This represents an important difference as compared to the doped B20 compounds, such as {\fcs} and {\mfs}, where the helical state under similar conditions is not recovered when approaching zero field. Further, the low-temperature skyrmion phase forms between the conical state and the field-polarised state highly hysteretic phase boundaries.  

The magnetic phase diagram inferred from combined FC/FSU and FC/FSD measurements are shown in Fig.\,\ref{figureS4c}\,(B). In order to explore the relationship between the high-temperature and the low-temperature skyrmion phase the field-cooling was carried out at $B=29$\,mT across the high-temperature skyrmion phase. This allowed to super-cool the high-temperature skyrmion phase to low temperatures. For positive field values (upper half of the diagram), the high-temperature skyrmion phase persists as a metastable state down to the lowest temperatures studied. Under increasing field the transition line remains below the phase boundary of the conical phase until the high-temperature skyrmion phase reaches the region in which the the low-temperature skyrmion phase is observed for negative fields. Due to strong dependence of the correlation length on the magnetic field (see below), it is not possible to confidently resolve whether both these phases coexist or only one survives in this phase region.  For negative field values, the same phase diagram is observed as recorded in the HFC/FSD measurements (lower part of Fig.\,\ref{figureS4c}\,(B)). This underscores, that the reversal of the orientation of the applied field resets the magnetic state. 

\newpage

\clearpage
%%%%%%%%%%%%%%%%%%%%%%%%%%%%%%%%%%%%%%%%%%%%%
\newpage
\section{Further microscopic details}

\subsection{Temperature and field dependence of the scattering intensities}

\begin{figure}[t]
\begin{center}
\includegraphics[width=0.9\textwidth]{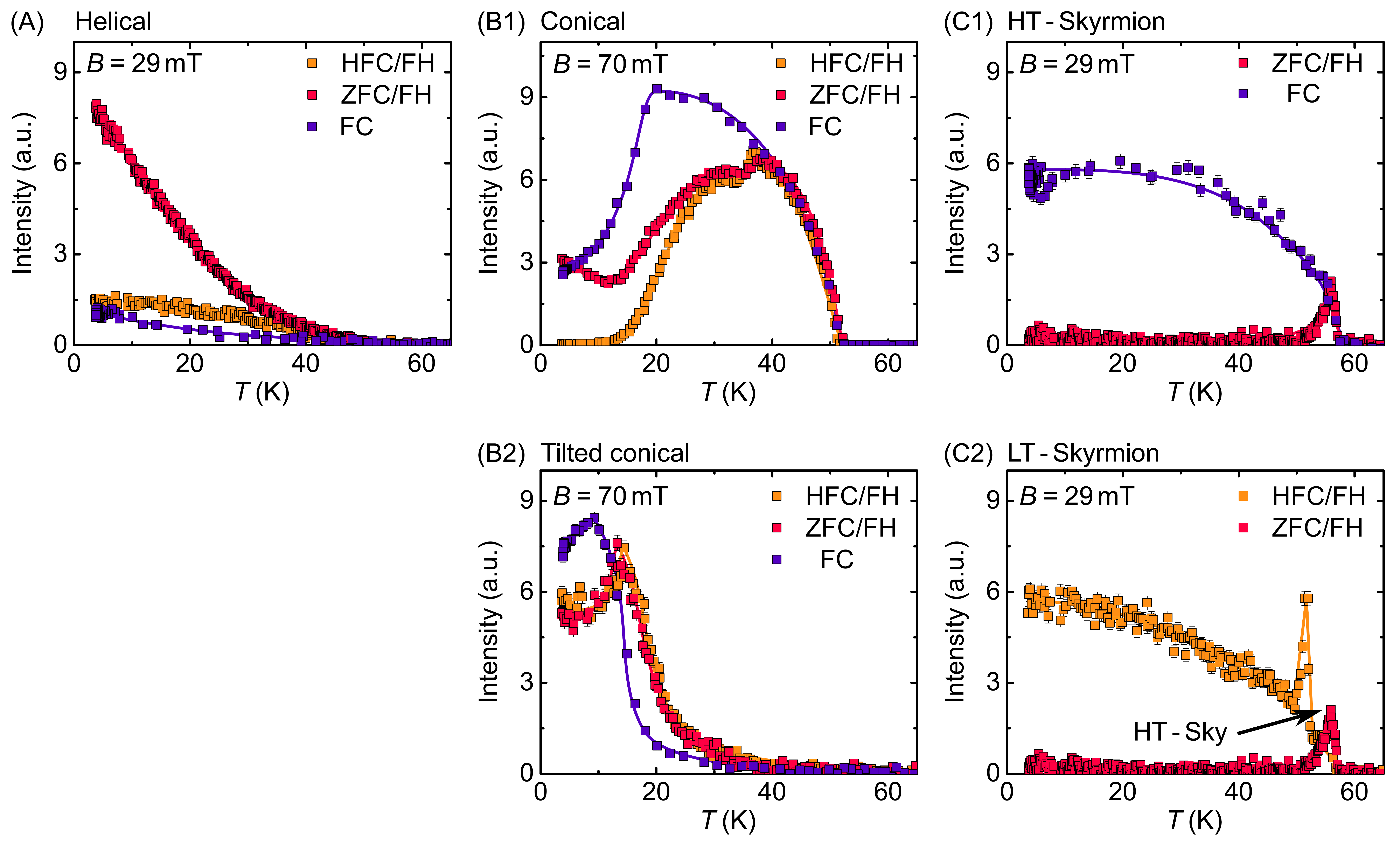}
\end{center}
\caption{
Comparison of the scattering intensities of the five different states as a function of temperature for the same characteristic magnetic field but different temperature versus field histories. 
\newline
%\textit{figure-S5a}
\label{figureS5a}
}
\end{figure}

\begin{figure}
\begin{center}
\includegraphics[width=0.9\textwidth]{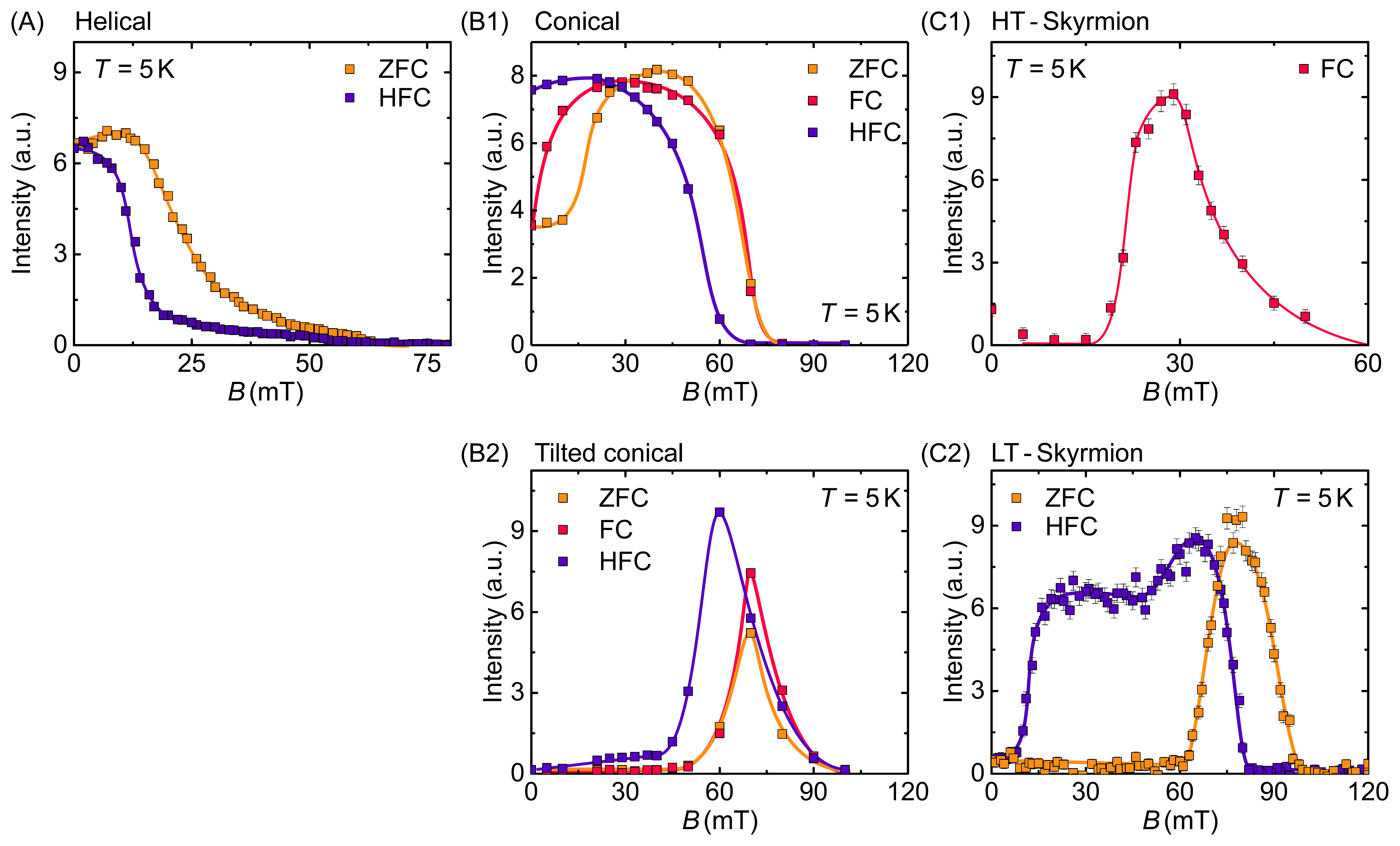}
\end{center}
\caption{
Comparison of the scattering intensities of the five different states as a function of magnetic field for the same temperature but different temperature versus field histories. 
\newline
%\textit{figure-S5b}
\label{figureS5b}
}
\end{figure}

The focus of our studies concerned the phase boundaries defining the magnetic phase diagram under various temperature versus field protocols. In general, both the phase boundaries and the intensities observed varied sensitively for different temperature and field histories. Typical data illustrating this aspect are shown in Fig.\,\ref{figureS5a}. All panels display data recorded as a function of temperature at the same fixed magnetic field as stated in each panel. Even though some of the phase boundaries do not change there are strong changes of the intensities. The same is also observed in field-sweeps as shown in Fig.\,\ref{figureS5b}.

\clearpage
%%%%%%%%%%%%%%%%%%%%%%%%%%%%%%%%%%%%%%%%%%%%%
\newpage

%%%%%%%%%%%%%%%%%%%%%%%%%%%%%%%%%%%%%%%%%%%%

\clearpage
%%%%%%%%%%%%%%%%%%%%%%%%%%%%%%%%%%%%%%%%%%%%%
\newpage

\subsection{Temperature and field dependence of correlation lengths}

\begin{figure}[t]
\begin{center}
\includegraphics[width=0.7\textwidth]{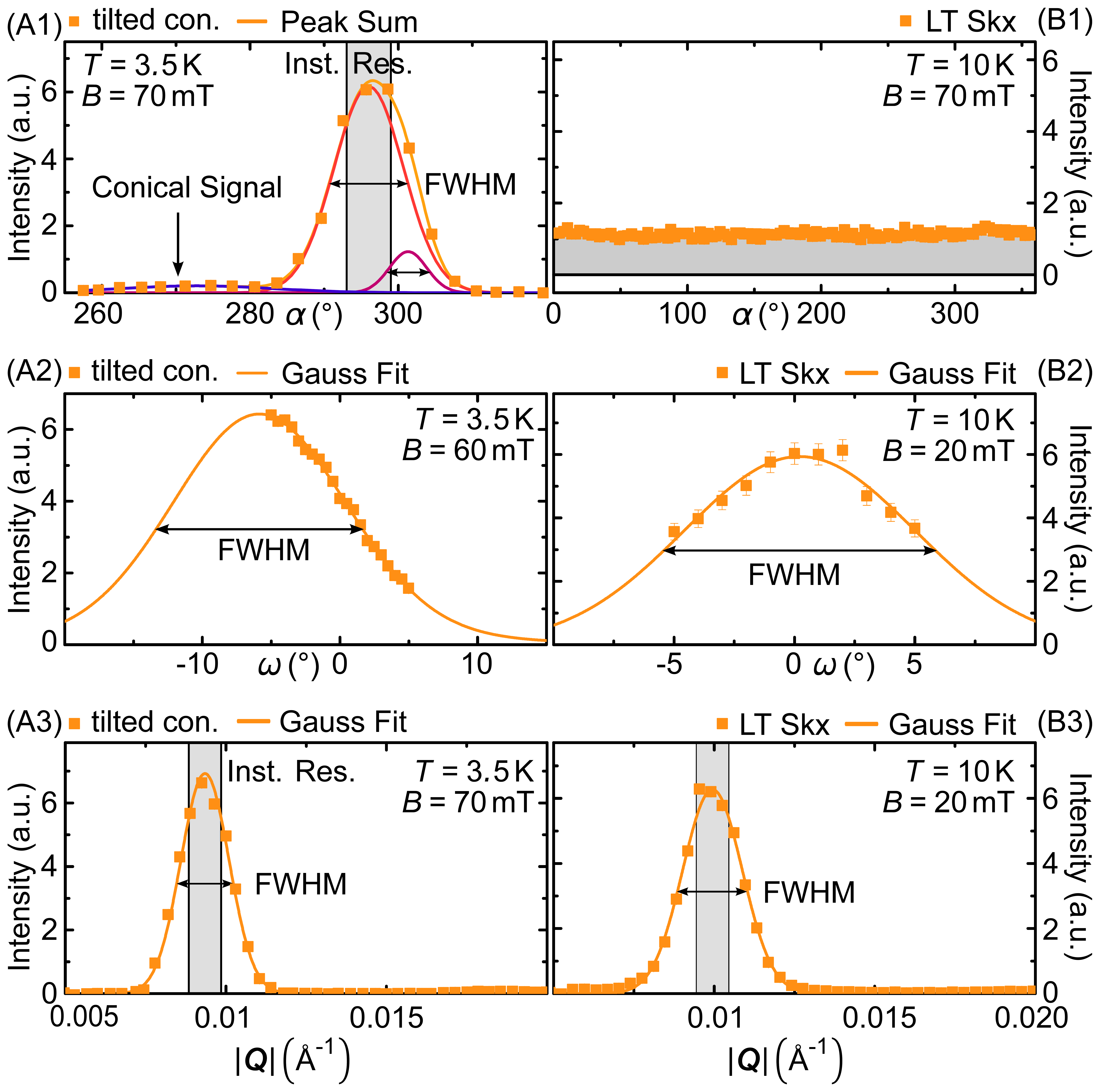}
\end{center}
\caption{
Typical variations of the scattering intensities of the tilted conical (til. con.) state and low-temperature skyrmion phase (LT-Sky) as a function of the azimuthal angle $\alpha$, the axis vertical (or horizontal) to the scattering plane, $\phi$, and the modulus of the scattering pattern, $\vert Q\vert$. In panels (A1), (A3) and (B3) the resolution limit is shown in gray shading. In panel (B1) the background is indicated as a line and the scattering by the sample in gray shading.
\newline
%\textit{figure-S6a}
\label{figureS6a}
}
\end{figure}

Typical variations of the scattering intensities of the tilted conical state and low-temperature skyrmion phase (LT-Sky) as a function of the azimuthal angle $\alpha$ within the scattering plane, the angle perpendicular to the scattering plane, $\phi$, and the modulus within the scattering pattern, $\vert Q\vert$, are shown in Fig.\,\ref{figureS6a}. As stated above, the resolution of the set up used in our study was $\Delta \alpha=\SI{6}{\degree}$, $\langle \Delta\vert Q\vert \rangle=\SI{0.0011}{\AA^{-1}}$, and $\Delta \phi =\SI{0.14}{\degree}$.  In panels (A1), (A3) and (B3) this resolution limit is shown in gray shading. In panel (B1) the background is indicated as a line and the scattering by the sample in gray shading. 

For the tilted conical state typical data may be summarised as shown in Figs.\,\ref{figureS6a}\,(A1) through (A3). The correlation length in die azimuthal direction (within the scattering plane) is close to the resolution limit and similar to the ones measured for the helical and conical state. The angular dependence (Fig.\,\ref{figureS6a}\,(A1)) seems to consist of two signals from domains separated by a couple of degrees. In contrast, the correlation length perpendicular to the scattering plane is very broad, greatly exceeding a typical FWHM\,$=\SI{3}{\degree}$ measured for the conical and helical state. The radial correlation length is again close to the resolution limit. For the low-temperature skyrmion phase, shown in Figs.\,\ref{figureS6a}\,(B1) through (B3), we do not observe any azimuthal dependence. As was the case for the tilted conical state, the correlation perpendicular to the scattering plane is very broad in comparison to other phases, while the radial correlation within the scattering plane is close to the resolution limit as shown in Figs.\,\ref{figureS6a}\,(B2) and (B3), respectively.

\begin{figure}[t]
\begin{center}
\includegraphics[width=0.99\textwidth]{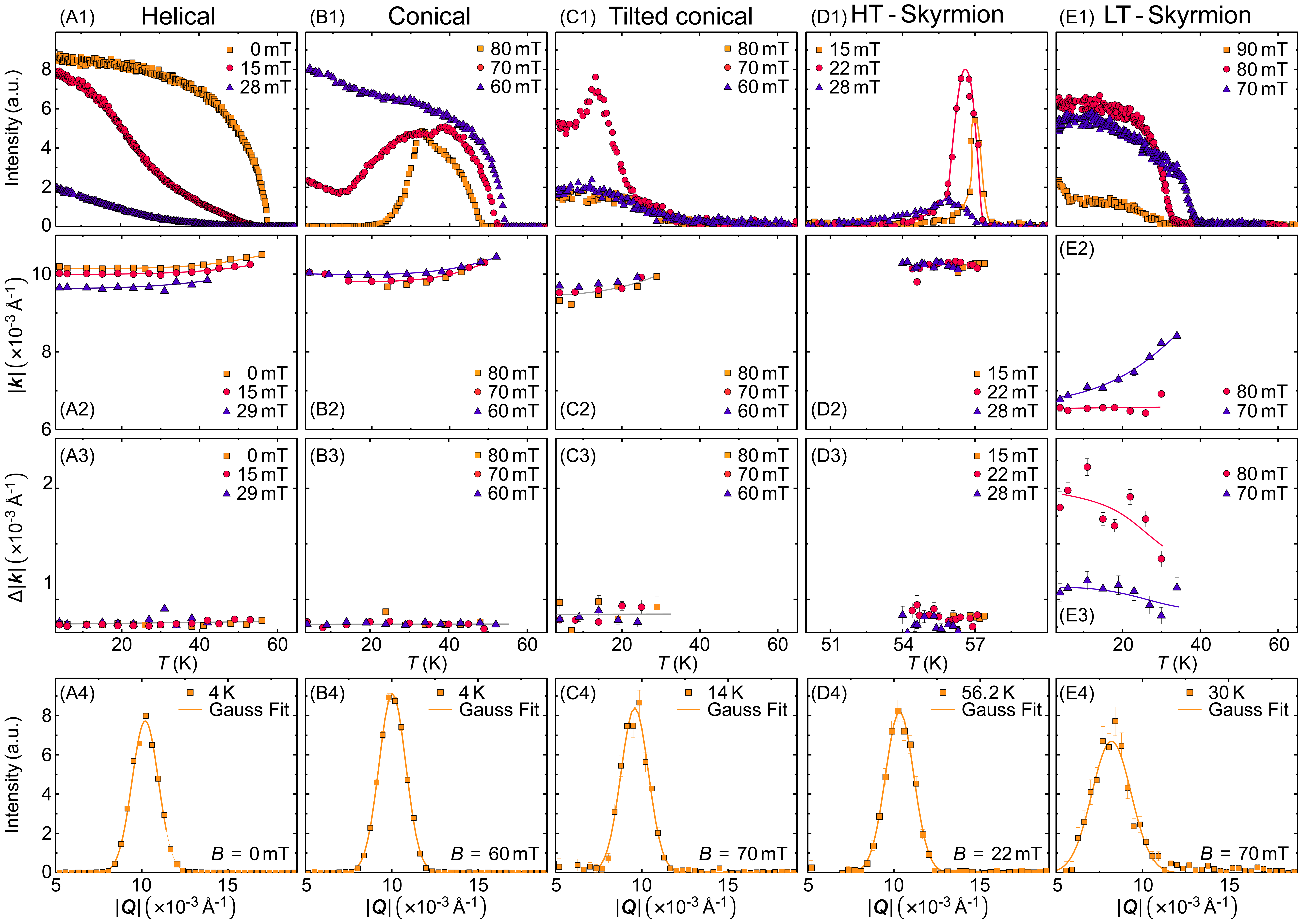}
\end{center}
\caption{ZFC/FH: 
Compilation of the scattering intensities, modulus of the modulation $\vert\bm{k}\vert$, variance of the modulation $\Delta\vert\bm{k}\vert$ and typical data used for the analysis of $\vert\bm{k}\vert$. Panels in column (A): Data of the helical modulation. Panels in column (B): Data of the conical state. Panels in column (C): Data of the tilted conical state. Panels in column (D): High temperature skyrmion phase. Panels in column (E): Low-temperature skyrmion phase.
\newline
%\textit{figure-S6b}
\label{figureS6b}
}
\end{figure}

\begin{figure}[t]
\begin{center}
\includegraphics[width=0.99\textwidth]{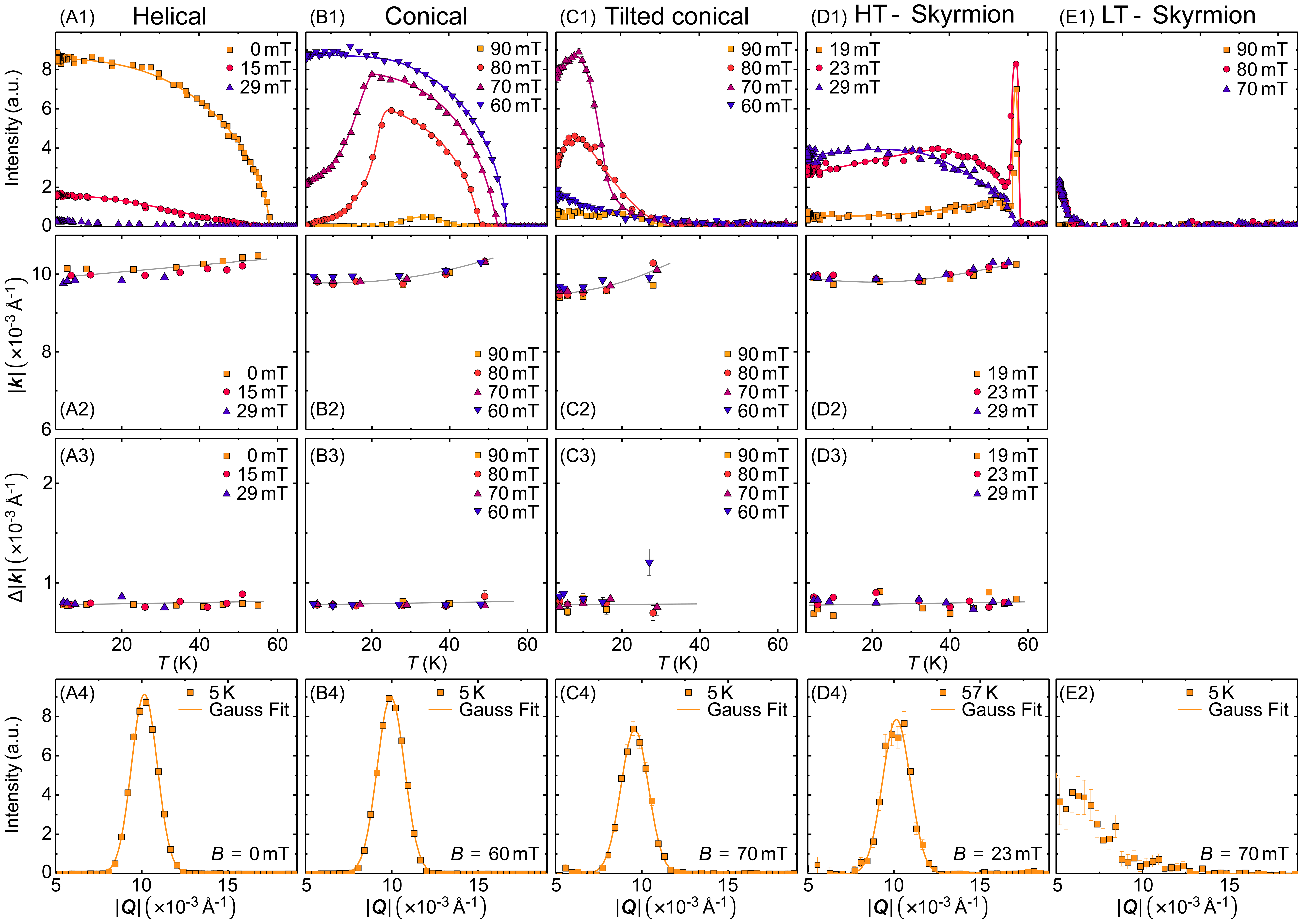}
\end{center}
\caption{FC:
Compilation of the scattering intensities, modulus of the modulation $\vert\bm{k}\vert$, variance of the modulation $\Delta\vert\bm{k}\vert$ and typical data used for the analysis of $\vert\bm{k}\vert$. Panels in column (A): Data of the helical modulation. Panels in column (B): Data of the conical state. Panels in column (C): Data of the tilted conical state. Panels in column (D): High temperature skyrmion phase. Panels in column (E): Low-temperature skyrmion phase.
\newline
%\textit{figure-S6c}
\label{figureS6c}
}
\end{figure}

\begin{figure}[t]
\begin{center}
\includegraphics[width=0.99\textwidth]{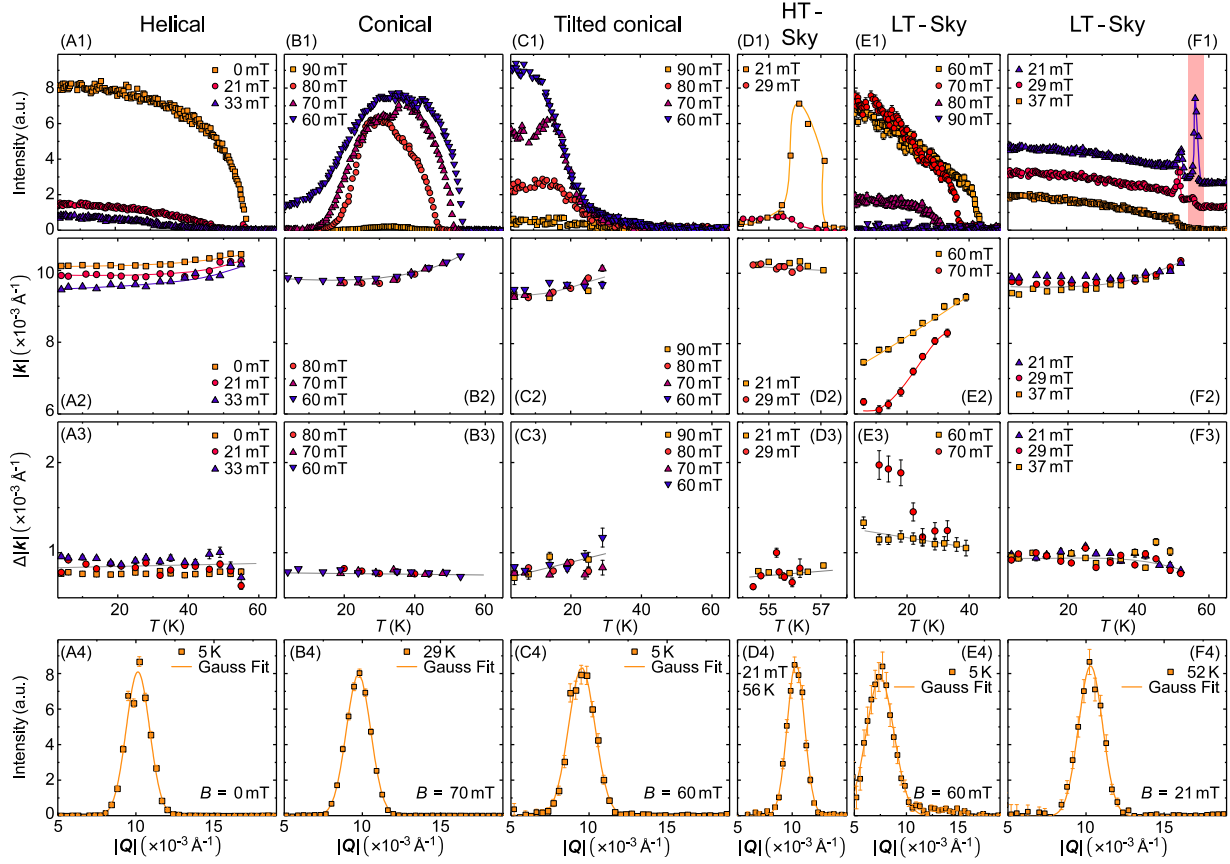}
\end{center}
\caption{HFC/FH:
Compilation of the scattering intensities, modulus of the modulation $\vert\bm{k}\vert$, variance of the modulation $\Delta\vert\bm{k}\vert$ and typical data used for the analysis of $\vert\bm{k}\vert$. Panels in column (A): Data of the helical modulation. Panels in column (B): Data of the conical state. Panels in column (C): Data of the tilted conical state. Panels in column (D): High-temperature skyrmion phase. Panels in column (E): Low-temperature skyrmion phase at high fields. Panels in column (F): Low-temperature skyrmion phase at low fields.
\newline
%\textit{figure-S6d}
\label{figureS6d}
}
\end{figure}

Shown in Figs.\,\ref{figureS6b}, \ref{figureS6c} and \ref{figureS6d} for ZFC/FH, FC and HFC/FH, respectively, are compilations of the scattering intensities, the modulus of the modulation $\vert \bm{k} \vert$, the variance of the modulation $\Delta\vert \bm{k}\vert$ as well as typical data used for the analysis of $\bm{k}$. The following general observations may be noted. As shown in the first row of each of the three figures the intensities depend sensitively on the measurement protocol used as already emphasised above. 

Inspection of the second row of the three figures reveals, that the modulus of the modulation for the different temperature versus field protocols and different magnetic states is always $\vert \bm{k} \vert \sim \SI{0.01}{\angstrom^{-1}}$. The only exception may be observed in the low-temperature skyrmion phase, as shown in Fig.\,\ref{figureS6b}\,(E2) and Fig.\,\ref{figureS6d}\,(E2). Here the modulus is small at low temperatures, $\vert k \vert \sim \SI{0.006}{\angstrom^{-1}}$, and increases with increasing temperature reaching $\vert \bm{k} \vert \sim \SI{0.01}{\angstrom^{-1}}$. Interestingly, for small fields the modulus of the modulation observed in the low-temperature skyrmion state is also $\vert \bm{k} \vert \sim \SI{0.01}{\angstrom^{-1}}$ without pronounced temperature dependence. The temperature and field dependence of $\vert \bm{k} \vert$ in the low-temperature skyrmion state compares with the tilt angle of the tilted conical state shown in Fig.\,\ref{figureS7a} below. For increasing field the tilt angle decreases characteristic of a decreasing strength of the anisotropy. In turn, this is consistent with the increase of the modulus shown in Fig.\,\ref{figureS6b}\,(E2) and Fig.\,\ref{figureS6d}\,(E2).

Last but not least, the third row of the three figures shows that the variance of the modulus, $\Delta\vert\bm{k}\vert$ is small around 
$0.8\cdot10^{-3}{\rm \AA}^{-1}$
for all magnetic states and all temperature versus field protocols. The only exception is shown in Figs.\,\ref{figureS6b}\,(E3) and Fig.\,\ref{figureS6d}\,(E3) where $\Delta\vert\bm{k}\vert$ is as large as 
$2\cdot 10^{-3}{\rm \AA}^{-1}$. 
This corresponds precisely to the situation when the modulus of the low-temperature skyrmion phase is small at low temperatures, increasing strongly with increasing temperature, cf. Fig.\,\ref{figureS6b}\,(E2) and Fig.\,\ref{figureS6d}\,(E2). As explained in the main text, the competing mechanisms of stabilisation and morphology, notably thermal fluctuations versus magnetic anisotropies and hexagonal skyrmion lattice versus square skyrmion lattice, where the latter has a reduced modulus, are in excellent agreement with the increased variance of the modulus and the lack of azimuthal dependence of the intensity observed experimentally. Taken together, these suggests for the low-temperature skyrmion phase a glassy, frustrated morphology of thermodynamically stable skyrmions.

\clearpage
%%%%%%%%%%%%%%%%%%%%%%%%%%%%%%%%%%%%%%%%%%%%%
\newpage

\clearpage
%%%%%%%%%%%%%%%%%%%%%%%%%%%%%%%%%%%%%%%%%%%%%
\newpage

%%%%%%%%%%%%%%%%%%%%%%%%%%%%%%%%%%%%%%%%%%%%

\clearpage
%%%%%%%%%%%%%%%%%%%%%%%%%%%%%%%%%%%%%%%%%%%%%
\newpage
\section{Miscelleaneous}

\subsection{Temperature dependence of the tilted conical state}

\begin{figure}[t]
\begin{center}
\includegraphics[width=0.5\textwidth]{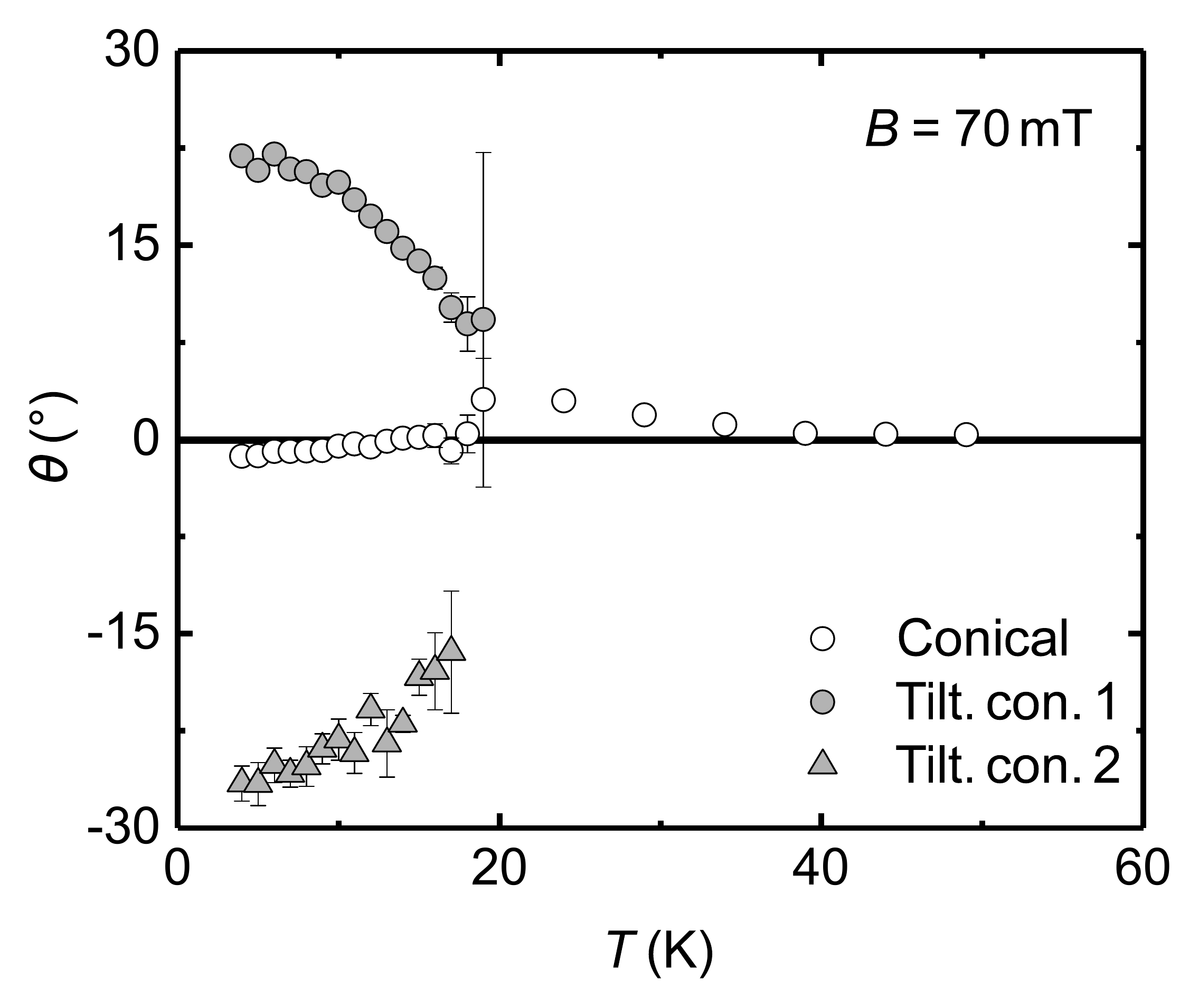}
\end{center}
\caption{
Temperature dependence of the tilt angle $\omega$ of the tilted conical state under HFC/FH. (A) Tilt angle $\alpha$ of the conical phase and tilted conical phase observed in HFC/FH measurements at 70\,mT.  
\newline
%\textit{figure-S7a}
\label{figureS7a}
}
\end{figure}

The temperature dependence of the tilt angle $\theta$ of the tilted conical state under HFC/FH as recorded in a magnetic field of 70\,mT is shown in Fig.\,\ref{figureS7a}. Measurements for this field value allowed to track the tilt angle as a function of temperature. However, the tilted conical phase and the conical phase were found to be in coexistence for the field value chosen. With increasing temperature the tilt angle decreases monotonically, highlighting the temperature induced reduction of the anisotropy term. The tilted conical phase vanishes above a critical temperature of 20\,K almost continuously, with a small discontinuity of the second domain. Shown in Fig.\,\ref{figureS7a}\,(B) is the width of the tilt angle of the tilted conical order as a function of temperature.  

\clearpage
%%%%%%%%%%%%%%%%%%%%%%%%%%%%%%%%%%%%%%%%%%%%%
\newpage

\clearpage
%%%%%%%%%%%%%%%%%%%%%%%%%%%%%%%%%%%%%%%%%%%%%
\newpage

\subsection{Modulus of the supercooled high-temperature skyrmion phase}

\begin{figure}[t]
\begin{center}
\includegraphics[width=0.5\textwidth]{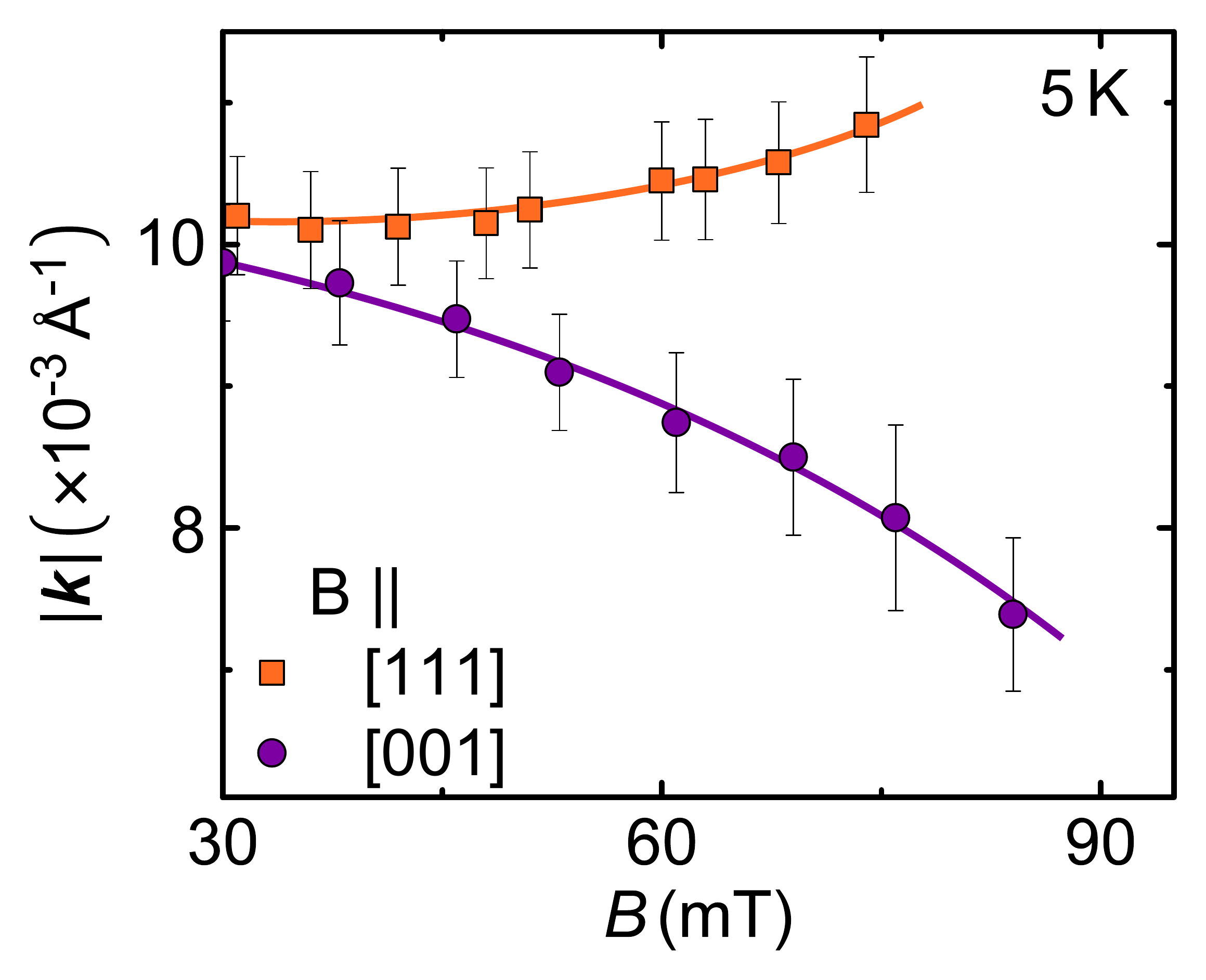}
\end{center}
\caption{
Comparison the magnetic field dependence of the modulus, $\vert \bm{k} \vert$, of the modulation of the high-temperature skyrmion phase at $\sim 5\,{\rm K}$ after field-cooling at a finite field of $\sim 29\,{\rm mT}$. For field parallel to the \hkl <100> a strong reduction is observed as shown in Fig.\,3 of the main text. In contrast, for magnetic field parallel \hkl <111> the modulus, $\vert \bm{k} \vert$, increases weakly. The difference illustrates the presence of an additional magnetic anisotropy for the \hkl <100> axis, that increases in strength with increasing field.   
\newline
%\textit{figure-S7b}
\label{figureS7b}
}
\end{figure}

A comparison of the magnetic field dependence of the modulus, $\vert \bm{k} \vert$, of the high-temperature skyrmion state at $\sim 5\,{\rm K}$ after field-cooling at a finite field of $\sim 29\,{\rm mT}$ is shown in Fig.\,\ref{figureS7b}. For field parallel $\langle 100 \rangle$ a strong reduction is observed as shown in Fig.\,3 in the main text. The reduction of $\vert \bm{k} \vert$ originates in the increase of the magnetic anisotropy with increasing magnetisation, causing an increase of the anharmonicity of the modulation. To reduce the associated increase of energy due to the increased gradients the modulus $\vert \bm{k} \vert$ decreases, i.e., the modulation length increases in order to reduce the gradients. Performing, in contrast, the same measurements for the same temperature versus magnetic field history but magnetic field parallel $\langle 111 \rangle$, we find that the modulus, $\vert \bm{k} \vert$, increases weakly. This underscores the importance of the $\langle 100 \rangle$ axis for the formation of the low-temperature skyrmion phase.

\clearpage
%%%%%%%%%%%%%%%%%%%%%%%%%%%%%%%%%%%%%%%%%%%%%
\newpage

\clearpage
%%%%%%%%%%%%%%%%%%%%%%%%%%%%%%%%%%%%%%%%%%%%%
\newpage

\subsection{Magnetic field sweep after ZFC and FC at an intermediate temperature}

\begin{figure}[t]
\begin{center}
\includegraphics[width=0.95\textwidth]{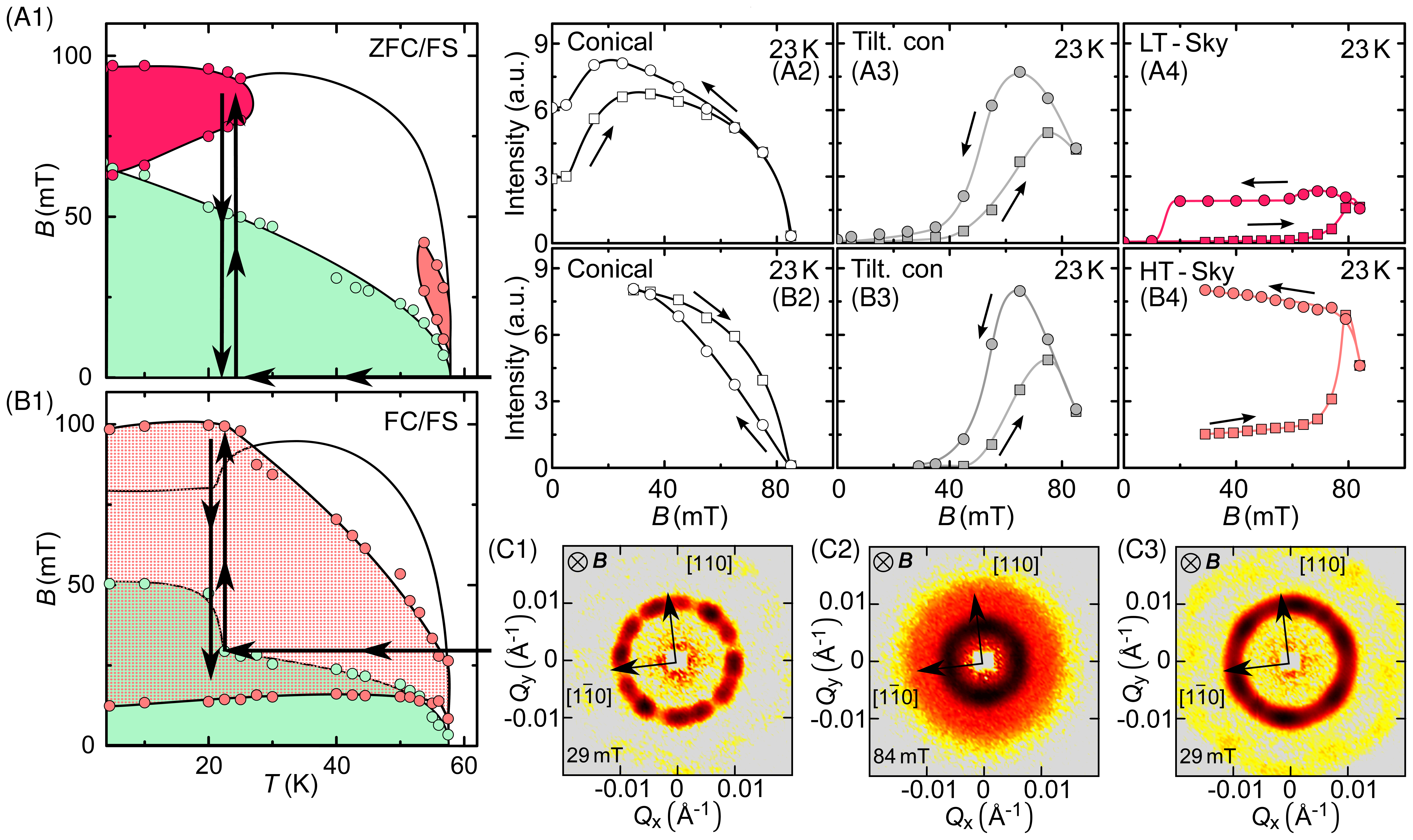}
\end{center}
\caption{
Evidence for the clear delineation of the mechanism of formation and morphology of the high-temperature skyrmion phase and the new low-temperature skyrmion phase as inferred from magnetic field scans. (A1) Magnetic phase diagram after zero-field cooling and a field-sweep up (ZFC/FSU). Red arrows indicate the specific temperature versus field trajectory of interest here. (A2) through (A4): For increasing field the conical intensity increases and vanishes. Under decreasing field the conical intensity is increased. The tilted conical state emerges at highest fields; it emerges also under decreasing field but vanishes. The ring of intensity characteristic of the low-temperature skyrmion state emerges under increasing field at the border to the field-polarised state. Under decreasing field the ring of intensity emerges and persists down to the transition to the helical state. (B1) Magnetic phase diagram after field cooling and a field-sweep up (FC/FSU). Red arrows indicate the specific temperature versus field trajectory of interest here. (B2) through (B4): The intensity of the conical state vanishes at large fields. Under decreasing fields it is \textit{lower} than for increasing field, suggesting presence of a competing ground state. The intensity of the tilted conical state is reminiscent to panel (A4). Most importantly, the sixfold intensity pattern of the high-temperature skyrmion state increases with increasing field. It increases even further upon decreasing field, in stark contrast with the  expectations of a metatstable state.
\newline
%\textit{figure-S8}
\label{figureS8}
}
\end{figure}

Shown in Fig.\,\ref{figureS8} is a comparison of the magnetic field dependence observed in field sweeps under different starting conditions, providing a clear delineation of the high-temperature and low-temperature skyrmion phases.  Shown in Fig.\,\ref{figureS8}\,(A1) is the magnetic phase diagram determined for ZFC/FSU. Of particular interest in the following is the behaviour when ZFC down to 23\,K, just above the temperature regime of the low-temperature skyrmion phase. A field-sweep of increasing field which terminates \textit{before} entering the field-polarised (ferromagnetic) state followed by a field-sweep back down to zero field reveals the behaviour shown in Figs.\,\ref{figureS8}\,(A2) through (A4). 

As shown in Fig.\,\ref{figureS8}\,(A2), the initial intensity due to the helical state observed at zero field increases before the state becomes conical around \SI{\sim20}{\milli\tesla}. It decreases again when approaching the field-polarised state above \SI{\sim80}{\milli\tesla}. Under decreasing field the intensity of the conical state is larger than for increasing field, indicating improved magnetic order. This compares with the field dependence of the tilted conical state, shown in Fig.\,\ref{figureS8}\,(A3). The intensity emerges above \SI{\sim50}{\milli\tesla} and decreases down to \SI{\sim90}{\milli\tesla}, the largest field measured. Under decreasing field the intensity increases again and is larger before it vanishes below \SI{\sim40}{\milli\tesla}. This establishes, that the conditions for stabilising the low-temperature skyrmion phase, notably a magnetic anisotropy that increases under increasing field, are barely met. Indeed, the ring of intensity as the key signature of the low-temperature skyrmion phase, emerges also under increasing field, but only around \SI{\sim75}{\milli\tesla} as shown in Fig.\,\ref{figureS8}\,(A4). Under decreasing field this ring of intensity remains stable down to \SI{\sim15}{\milli\tesla}, the transition to the helical state. It is essential to emphasise, that the ring of intensity as the signature of the low-temperature skyrmion phase during this field sweep does not exhibit any azimuthal dependence suggesting a lattice formation.

Shown in Fig.\,\ref{figureS8}\,(B1) is the magnetic phase diagram observed for combined FC/FSU and FC/FSD when field-cooling at 29\,mT across the high-temperature skyrmion phase. The behaviour observed in a field sweep after field-cooling up to field slightly smaller than the transition field to the field-polarised state is summarised in Figs.\,\ref{figureS8}\,(B2) through (B4) and panels (C1) through (C3). The intensity of the conical state, shown in Fig.\,\ref{figureS8}\,(B2), decreases with increasing magnetic field and vanishes above \SI{\sim85}{\milli\tesla}. A first surprise is the variation of the conical intensity under decreasing field, which remains \textit{below} the value observed under increasing field. This suggests the presence of a different ground state that is energetically advantageous. In contrast, the field dependence of the tilted conical state shown in Fig.\,\ref{figureS8}\,(B3) is reminiscent of the behaviour observed after ZFC (cf. Fig.\,\ref{figureS8}\,(A4)), where the scan stops again before reaching the field-polarised state. 

A major difference as compared to the behaviour observed for ZFC concerns the super-cooled high-temperature skyrmion phase, shown in Fig.\,\ref{figureS8}\,(B4), (C1), (C2) and (C3). For increasing field the intensity increases and reaches a maximum just below the transition to the field-polarised state. However, when decreasing field again the intensity {\it continues to grow} reaching a very large value before collapsing \SI{\sim15}{\milli\tesla}, below which the helimagnetic state forms. Moreover, the diffraction pattern of the FC high-temperature skyrmion state, shown in Fig.\,\ref{figureS8}\,(C1), displays the azimuthal dependence of two domain populations of the hexagonal skyrmion lattice. With increasing field the modulus of the pattern decreases strongly, forming a nearly uniform ring at large fields as shown in Fig.\,\ref{figureS8}\,(C2). Finally, under decreasing field, a broadened sixfold pattern stabilises, characteristic of the high-temperature skyrmion state as shown in Fig.\,\ref{figureS8}\,(C3). 

The behaviour observed here highlights (i) the presence of two different mechanism stabilising the skyrmion phase, and (ii) a distinct difference of the morphology of the high-temperature and low-temperature skyrmion phases. Regarding the mechanisms stabilising the phases these are, one the hand, the mode coupling term in combination with thermal fluctuations and, on the other hand, the magnetic anisotropies arising from the crystallographic symmetry. For the temperature of 23\,K at which the data shown in Fig.\,\ref{figureS8} was recorded the effects of both mechanisms are reduced and conspire. Namely, as shown in Fig.\,\ref{figureS7a} in terms of the reduction of the tilt angle of the tilted conical state, the strength of the magnetic anisotropy decreases with increasing temperature and is almost zero. Likewise, the effects of thermal fluctuations must also be strongly reduced as the temperature is reduced by over $60\,\%$ as compared to the temperature range around \SI{\sim60}{\kelvin}, where the high-temperature skyrmion phase forms as a thermodynamically stable state. 

Strong evidence for the inherent difference of the morphology of the high-temperature and low-temperature skyrmion phases represents the pronounced six-fold azimuthal intensity variation that stabilises in the field cycle as shown in Figs.\,\ref{figureS8}\,(C1) through (C3).

%\clearpage
%%%%%%%%%%%%%%%%%%%%%%%%%%%%%%%%%%%%%%%%%%%%%
%\newpage

\clearpage
%%%%%%%%%%%%%%%%%%%%%%%%%%%%%%%%%%%%%%%%%%%%%
\newpage

\bibliography{LT-Skx}